\documentclass[preprint]{elsarticle}
\usepackage{lineno}
\modulolinenumbers[5]

\usepackage[utf8]{inputenc}
\usepackage[english]{babel}
\usepackage{bbold}
\usepackage[T1]{fontenc}
\usepackage{amsmath}
\usepackage{amsfonts}
\usepackage{amssymb}
\usepackage{amsthm}
\usepackage{graphicx}
\usepackage{stmaryrd}
\usepackage{subfig}
\usepackage[left=2cm,right=2cm,top=2cm,bottom=2cm]{geometry}
\usepackage[dvipsnames]{xcolor}
\usepackage{answers}
\usepackage{enumitem}

\newcommand{\cad}{\textit{ie.} }
\newcommand{\lsf}{limit-state function }
\newcommand{\E}[1]{\operatorname{E}\left[ #1 \right]}
\newcommand{\var}[1]{\operatorname{var}\left[ #1 \right]}
\newcommand{\cov}[2]{\operatorname{cov}\left[ #1, #2 \right]}

\newcommand{\proba}[1]{\operatorname{P}\left[ #1 \right]}

\DeclareMathAlphabet{\mathonebb}{U}{bbold}{m}{n}
\newcommand{\one}{\ensuremath{\mathonebb{1}}}
\newcommand{\X}{\ensuremath{\mathbf{X}}}
\newcommand{\x}{\ensuremath{\mathbf{x}}}
\newcommand{\U}{\ensuremath{\mathbf{u}}}
\newcommand{\R}{\ensuremath{\mathbb{R}}}
\newcommand{\N}{\ensuremath{\mathbb{N}}}
\newcommand{\p}{\ensuremath{\widehat{p}}}

\newcommand{\eq}[1]{\begin{equation} \begin{array}{rcl} #1 \end{array} \end{equation}}

\newcommand{\tmk}{T_{m+k}}
\newcommand{\tmt}{T_{M_t}}
\newcommand{\tmtun}{T_{M_t+1}}

\newcommand{\qmk}{\ensuremath{q_{m+k}}}
\newcommand{\qcl}{ \widehat{q}}
\newcommand{\tend}[1]{\overset{\mathcal{L}}{\underset{#1 \rightarrow \infty}{\longrightarrow}}}
\newcommand{\conv}[1]{\underset{#1 \rightarrow \infty}{\longrightarrow}}
\newtheorem{theo}{Theorem}
\newtheorem{algo}{Algorithm}
\newtheorem{lemme}{Lemma}
\newtheorem{coro}{Corollary}
\newtheorem{propo}{Proposition}
\newtheorem{rem}{Remark}

\newcommand{\za}{Z_{1-\alpha/2}}
\newcommand{\mx}{\text{max}}
\newcommand{\mn}{\text{min}}
\newcommand{\Niter}{N_\text{iter}}
\newcommand{\tpar}{t_\text{par}}
\newcommand{\tmc}{t_\text{MC}}
\newcommand{\tsmc}{t_\text{MS}}
\newcommand{\loi}{\overset{\mathcal{L}}{\sim}}
\newcommand{\ncall}{N_\text{call}}

\Newassociation{proofprop}{demprop}{ann}

\Newassociation{prooftheo}{demtheo}{ann}

\Newassociation{prooflemme}{demlemme}{ann}

\Newassociation{proofcoro}{demcoro}{ann}

\begin{document}

\begin{frontmatter}

\title{Moving Particles: a parallel optimal Multilevel Splitting method with application in quantiles estimation and meta-model based algorithms}

\author[mymainaddress,mysecondaryaddress]{Clément WALTER\corref{mycorrespondingauthor}}
\cortext[mycorrespondingauthor]{Corresponding author. Tel.: +33 1 69 26 66 75; Cel.: +33 6 65 16 53 97}
\ead{clement.walter@cea.fr}

%\author[mysecondaryaddress]{Josselin GARNIER}
%\ead{garnier@math.univ-paris-diderot.fr}
\address[mymainaddress]{CEA, DAM, DIF, F-91297 Arpajon, France}%
\address[mysecondaryaddress]{Université Paris Diderot, 5 rue Thomas Mann, 75013 Paris, France}%
\begin{abstract}
Considering the issue of estimating \emph{small} probabilities $p$, \textit{ie.} measuring a \emph{rare} domain $F = \{ \x \mid g(\x) > q \}$ with respect to the distribution of a random vector $\X$, Multilevel Splitting strategies (also called \textit{Subset Simulation}) aim at writing $F$ as an intersection of \emph{less rare} events (nested subsets) such that their measures are conditionally easily computable. However the definition of an \emph{appropriate} sequence of nested subsets remains an open issue.

We introduce here a new approach to Multilevel Splitting methods in terms of a move of particles in the input space. This allows us to derive two main results: (1) the number of samples required to get a realisation of $\X$ in $F$ is drastically reduced, following a Poisson law with parameter $\log 1/p$ (to be compared with $1/p$ for naive Monte-Carlo); and (2) we get a parallel optimal Multilevel Splitting algorithm where there is indeed no subset to define any more.

We also apply result (1) in quantile estimation producing a new parallel algorithm and derive a new strategy for the construction of first Design Of Experiments in meta-model based algorithms.
\end{abstract}
\begin{keyword}
Rare event simulation \sep MCMC \sep Sequential Monte-Carlo \sep Subset Simulation
\end{keyword}

\end{frontmatter}

\linenumbers

\Opensolutionfile{ann}[demo]

\section{Introduction}

\paragraph{Context}

Extreme events simulation and quantification come from the need to insure that undesirable events will not appear. Typically such events are failure of industrial critical systems, \textit{ie.} systems for which failure is regarded as a massive catastrophic situation, found in sectors like nuclear safety, aerospace, etc. In this context one could either want to estimate a probability of failure or to define a threshold to insure security with a given confidence. Usually the system is a "black box" whose output determines safety/failure domains.

Formally, let $\mathbf{X}$ be a random vector with values in $\mathbb{R}^d$ and $g$ be a measurable function from $\mathbb{R}^d$ to $\mathbb{R}$ defining the failure domain $F = \{ \mathbf{x} \in \mathbb{R}^d \, | \, \operatorname{g}(\mathbf{x}) > q \}$, we seek to measure $F$ with respect to the distribution of $\mathbf{X}$: $\proba{\X \in F} = \mu^X(F) = p$ or to find back $q$ given $p$.

Two reasons why this calculation is not obvious is the order of magnitude of the probability (say $p<10^{-5}$) and the computational time of $g$ output, from couples of hours to several months. In this framework efficiency of the algorithm (precision, global computational time) with respect to good statistical properties of the estimator are of great interest.

Let us first introduce several techniques used so far. Then we will present our new algorithm and its applications in probability and quantile estimation.

\paragraph{Modified Monte-Carlo algorithms}

A comprehensive review of Monte-Carlo methods can be found in \cite{rubinstein2011simulation}. On the one hand Importance Sampling \cite{robert2004monte,asmussen2007stochastic} modifies the distribution of $\X$ to lower the variance of the naive Monte-Carlo estimator; unfortunately the search for an appropriate change of probability is not obvious. In particular it is known that the optimal change depends on the quantity of interest and is thus not directly available.

On the other hand Multilevel Splitting methods consider the failure domain as a finite intersection of nested subsets for which conditional probabilities are \emph{not too small} and so more easily computable: let $(F_k)_k$ be a finite sequence of nested subsets ($F_0 = \R^d$) such that $F = \bigcap \limits_k F_k$, one can write:
$$
\proba{g(\mathbf{X}) > q} = \mu^X(F) = \prod \limits_k \mu^X(F_k \mid F_{k-1})
$$
In these algorithms the two mains issues are the conditional sampling and the subsets definition. The idea of splitting an event $F = \{ g(\X) > q\}$ with a sequence of $(q_m)_m$ such as $F$ can be written as an intersection of nested subsets appeared in the mid 1950's (from Kahn\&Harris \cite{kahn1951estimation} and Rosenbluth\&Rosenbluth \cite{rosenbluth2004monte}). Then Au\&Beck \cite{au2001estimation} brought it to rare event estimation. An in-depth review of these techniques can be found in Glasserman \textit{et al.} \cite{glasserman1999multilevel}. This algorithm was further improved by Cérou \textit{et al.} \cite{cerou2006genetic} linking it with Feynman-Kac formulae and Cérou and Guyader \cite{cerou2007adaptive} who proposed a method to adaptively select the conditional probabilities. Concerning the conditional simulations, Del Moral \textit{et al.} \cite{del2006sequential} introduced reversible transition kernels. Finally Cérou \textit{et al.} \cite{cerou2012sequential} showed that these algorithms were optimal when all conditional probabilities were equal and that an adaptive choice of levels leads to bias in the estimators. Recently Guyader \textit{et al.} \cite{guyader2011simulation} proposed a limit case where the conditional probabilities are fixed to $1-1/N$, given $N$ the size of the working population. While they showed that this choice is optimal in terms of computational efficiency, it also disables parallel computation possibility and eventually makes this algorithm longer in practice if multicore computers are available.

\paragraph{Meta-model based algorithms}

As modified Monte-Carlo methods seen above still require an important number of samples and do not allow for full parallelisation, meta-model based algorithms propose to spend the computational budget in fitting a surrogate model to the expensive-to-evaluate function $g$ and then to use it instead of the \emph{true} function to compute probability estimation with usual methods \cite{echard2011ak,bourinet2011assessing,dubourg2011metamodel}. Thus theses strategies highly depend on the \emph{quality} of the Design of Experiments (DoE) and especially on their ability to predict the boundary between safety and failure domains, \textit{ie.} to explore the input space close to the boundary. While space-filling strategies recommend to sample uniformly in the input space (see \cite{santner2003design} or \cite{dubourg2011metamodel} chapter 2 for a review of these methods) and thus depend a lot on the dimension of the input space, DoE generated according to the distribution of $\X$ are unlikely to visit the failure domain because of the order of magnitude of the failure probability.

\paragraph{Effective computing time} As mentioned previously, the function of interest $g$ is assumed to be highly demanding in computational time and the number of calls to $g$, \cad the number of samples used for an estimator, is limited. In this context parallel algorithms (see for examples reference books \cite{kumar1994introduction} or \cite{dongarra2003sourcebook}) are of great interest. Basically they allow for generating samples for an estimator in a parallel way, \cad that one can get as many samples as available "computers" for the time of one. Thus, to increase the precision of an estimator one can simply use more computers, \cad multi-core processors, to get more samples without making the estimation longer. We will refer to the number of calls to the limit-state function made by one computer as the effective computing time of an algorithm.

\paragraph{Main results}
We introduce here a new approach to Multilevel Splitting in terms of a move of particles from an initial random state to the failure domain. This approach brings two main results: first the number of samples needed to get a realisation of $\X$ in the failure domain follows a Poisson law with parameter $\log 1/p$, this is to be compared with a classical Geometric law with parameter $p$ for naive Monte-Carlo; then we get the full parallelisation of the optimal sequential algorithm described by Guyader \textit{et al.} \cite{guyader2011simulation}, which turns it into the best Multilevel Splitting algorithm in terms of effective computing time, resolving the issues of choosing a sequence of $(q_m)_m$ or selecting a cut-off probability for the adaptive construction. This new point of view also allows us to propose a modified version of Guyader \textit{et al.} quantile estimator with a reduced bias and parallel computation.

In the context of meta-model based algorithms (that require DoE with points close to the boundary between safety and failure domains), we use this approach to get a first DoE embedding failing samples while limiting drastically the number of calls to the limit-state function, which depends linearly on the dimension:
$$N_{DoE} = d + 1 + N_\text{fail}  \log(1/p)$$
with $N_\text{DoE}$ the size of the first DoE, $d$ the dimension of the input space and $N_\text{fail}$ the final number of points in the failure domain.

\section{Getting into the failure domain}

\subsection{Introduction}

The idea of trying to go \emph{as fast as possible} into the failure domain comes from the need to get failing samples in several methods, from Importance Sampling \cite{au1999new} to meta-model based algorithm. In this latter case it is well noticed \cite{echard2011ak} that without a first DoE embedding failing samples, the learning of the failure domain is complicated and the final probability estimator rather poor. On the other hand it was tried to merge Multilevel Splitting methods and meta-model based algorithms to increase the precision of conditional probabilities estimation while making easier the learning of the failure domain \cite{bourinet2011assessing, li2012bayesian}. The paradox was that the final DoE was indeed far too dense in \emph{a posteriori} useless regions.

Then we came up with the idea of \emph{stopping} to try computing the probability estimation on-the-go but only keeping the \emph{moving} part of these algorithms. Finally, Guyader \textit{et al.} work \cite{guyader2011simulation} brings to us the theoretical framework to derive this algorithm.

The problem can be defined as follows: let $\mathbf{X}$ be a random variable with values in $\mathbb{R}^d$, $d \in \mathbb{N}^*$, $\mu^X$ its distribution and $g$ a measurable function from $\mathbb{R}^d$ to $\mathbb{R}$ such that the \textit{cdf} of $g(\X)$ is continuous.

We first introduce the algorithm in the ideal case, \cad the case where we know how to sample from any distribution when required, and we will then present two implementations to be used depending on the goal (probability and quantile estimation or building of first DoE).

\subsection{Ideal case}

In this section we consider that it is possible to sample from any distribution when required; thus it is said \emph{ideal}.
\subsubsection{Move of one particle} \label{move_one_part}

\begin{algo}[Move of one particle] \ \label{algo_atteinte_defaillance}
\begin{itemize}
\item $q_0 = -\infty$
\item Iteration over $m$:
	\begin{itemize}
	\item[+] Sample $\X \sim \mu^X(\, \cdot \, \mid g > q_m)$
	\item[+] Evaluate $g$: $q_{m+1} = g(\X)$
	\end{itemize}
\end{itemize}
\end{algo}
Thus, Algorithm \ref{algo_atteinte_defaillance} can be seen as a move of a particle from an initial random place along the levels of $g$ and the sequence of $(q_m)_m$ has indeed an interesting statistical behaviour. Let $F_g$ be the \textit{cdf} of $g(\X)$ and $\Lambda$ the integrated hazard function: $\Lambda(y) = -\log (1 - F_g(y) )$; the following results are based on \cite{guyader2011simulation}:

\begin{theo} \label{theo_algo_ppp}
The random variables $(T_m)_{m \geq 1} = (\Lambda(q_m))_{m \geq 1}$ are distributed as the successive arrival times of a Poisson Process with parameter 1.

\begin{prooftheo}
First $\Lambda$ is an increasing function from $\R$ to $\R_+$ and the sequence $(q_m)_m$ is increasing and so the sequence $(T_m)_m$ is.

It remains to show that it is a Poisson Process with parameter 1, which means by definition that inter-arrival times are independent and follow an exponential law with parameter 1. Considering $m \in \N$ (with the convention $q_0 = -\infty$ and $T_0 = 0$) we have:
$$
\begin{array}{rcl}
T_{m+1} - T_m & = & \Lambda(q_{m+1}) - \Lambda(q_m) = -\log ( \proba{g(\X) > q_{m+1}} ) + \log ( \proba{g(\X) > q_m }) \\[0.5cm]
& = & - \log \left( \dfrac{ \proba{g(\X) > q_{m+1}} }{ \proba{g(\X) > q_m} } \right) \\[0.5cm]
T_{m+1} - T_m & = & - \log \left( \proba{ g(\X) > q_{m+1} \mid g(\X) > q_m } \right)
\end{array}
$$
Let $\mathcal{F}_m$ be the $\sigma$-algebra generated by $(T_j)_{j \leq m}$ and $F_m$ be the \textit{cdf} of $g(\X)$ when the distribution of $\X$ is $\mu^X(\, \cdot \, \mid g > q_m)$. Knowing $\mathcal{F}_m$, $F_m$ is the \textit{cdf} of $q_{m+1}$ and thus $F_m(q_{m+1})$ follows a uniform law on $[0; 1]$. Finally we get:
\begin{equation}
\begin{array}{rcl}
\forall t \in \R_+, \proba{ T_{m+1} - T_m < t \mid \mathcal{F}_m} & = & \proba{ - \log(1-F_m(q_{m+1})) < t \mid \mathcal{F}_m} \\
& = & \proba { F_m(q_{m+1}) < 1 - \exp(-t)\mid \mathcal{F}_m} \\
\forall t \in \R_+, \proba{ T_{m+1} - T_m < t\mid \mathcal{F}_m} & = & 1 - \exp(-t)
\end{array}
\end{equation}
Thus the inter-arrival times are independent and follow an exponential law with parameter 1 and $(T_m)_m$ is a Poisson Process.
\end{prooftheo}
\end{theo}

\begin{coro} \label{coro_nbr_mutations_1_particule}
The number of moves needed to get a realisation of $\X$ in a domain $F = \{ \mathbf{x} \in \mathbb{R}^d \, | \, \operatorname{g}(\mathbf{x}) > q \}$, $q \in \mathbb{R}$, with probability measure $\mu^X(F) = p$ follows a Poisson law with parameter $\log 1/p$.

\begin{proofcoro}
In this context we focus on the time $t = -\log(1 - F(q)) = -\log p$. The number of events before $t$ equals the number of moves to get the first realisation of $\X$ in $F$ and thus follows a Poisson law with parameter $t = \log 1/p$.
\end{proofcoro}
\end{coro}
Proofs of these results are given in the appendix.

\subsubsection{Move of $N$ particles} \label{move_N_part}
We now detail how to move a bunch of particles all together and show that it is indeed exactly the same as moving particles all separately. The following algorithm is almost the one proposed by Guyader \textit{et al.} \cite{guyader2011simulation} in the context of Multilevel Splitting algorithms, but presented in the scope of moving particles; in particular we do not sort the $(q_m^j)_j$ at a given iteration $m$ to keep the particles tractable.

\begin{algo}[Move of $N \in \N^*$ particles] \ \label{algo_atteinte_defaillance_par}
\begin{itemize}
\item $\mathbf{q}_0 = (q_0^1, \cdots, q_0^N) = (-\infty, \cdots, -\infty)$
\item Iteration over $m$:
	\begin{itemize}
	\item[+] $\mathbf{q}_{m+1} = \mathbf{q}_m$
	\item[+] $i_m = \underset{i}{\operatorname{argmin}}(q_m^i)$
	\item[+] Sample $\X_{i_m} \sim \mu^X(\, \cdot \, \mid g > q_m^{i_m})$
	\item[+] Evaluate $g$: $q_{m+1}^{i_m} = g(\X_{i_m})$
	\end{itemize}
\end{itemize}
\end{algo}
At time $m = N$ $(\X_1, \cdots \X_N)$ is an iid sample of the distribution $\mu^X$. Afterwards the move of $N$ particles all together is made sequentially by moving at each step the last one. Once again we study the properties of the sequence of $(\mathbf{q}_m)_m$. With the same notations and hypothesis as in section \ref{move_one_part}, we have (proofs in the appendix):

\begin{theo}
The random variables $(T_m)_{m \geq N} = (\Lambda(q_m^{i_m}))_{m \geq N}$ are distributed as the successive arrival times of a marked Poisson Process with parameter $N$

\begin{prooftheo}
There is only one move per iteration and so each particle is indeed moved according to Algorithm \ref{algo_atteinte_defaillance}, \cad conditionally to its current position and the times only depend on the position. Given $i_m$ the index of the particle realising the $m^{th}$ minimum: $\forall m \geq N, i_m = \underset{i}{\operatorname{argmin}} \, q_m^i$ (if several particles achieve the minimum $i_m$ is chosen uniformly amongst the indices), we build:
$$T_m \longmapsto (T_m,i_m)$$
which is a marked Point Process in $\mathbb{R^*_+} \times \mathbb{N^*}$ (see \cite{last1995marked} for more details about marked Point Processes). The processes on each mark are independent Poisson Process with parameter 1 and algorithm \ref{algo_atteinte_defaillance_par} is finally a marked Poisson Process with parameter $\sum 1 = N$
\end{prooftheo}
\end{theo}

\begin{coro} \label{coro_nbr_mutations_N_particules}
The number of moves needed to get $N$ realisations of $\X$ in a domain $F = \{ \x \in \R^d \mid g(\x) > q \}$, $q \in \R$, with probability measure $\mu^X(F) = p$ follows a Poisson law with parameter $N \log 1/p$.

\begin{proofcoro}
The Poisson process is marked and so the particles move indeed independently. As expressed by Corollary \ref{coro_nbr_mutations_1_particule} each particle has to move $M \loi \mathcal{P}(\log 1/p)$ times. Finally the total number of moves is the sum of the number of moves of each particle and thus follows a Poisson law with parameter $N \log 1/p$.
\end{proofcoro}
\end{coro}

\begin{center}
\begin{figure}[!ht]
\begin{minipage}[c]{0.5\textwidth}
\subfloat[Three realisations of algorithm \ref{algo_atteinte_defaillance}]{\includegraphics[width=\textwidth]{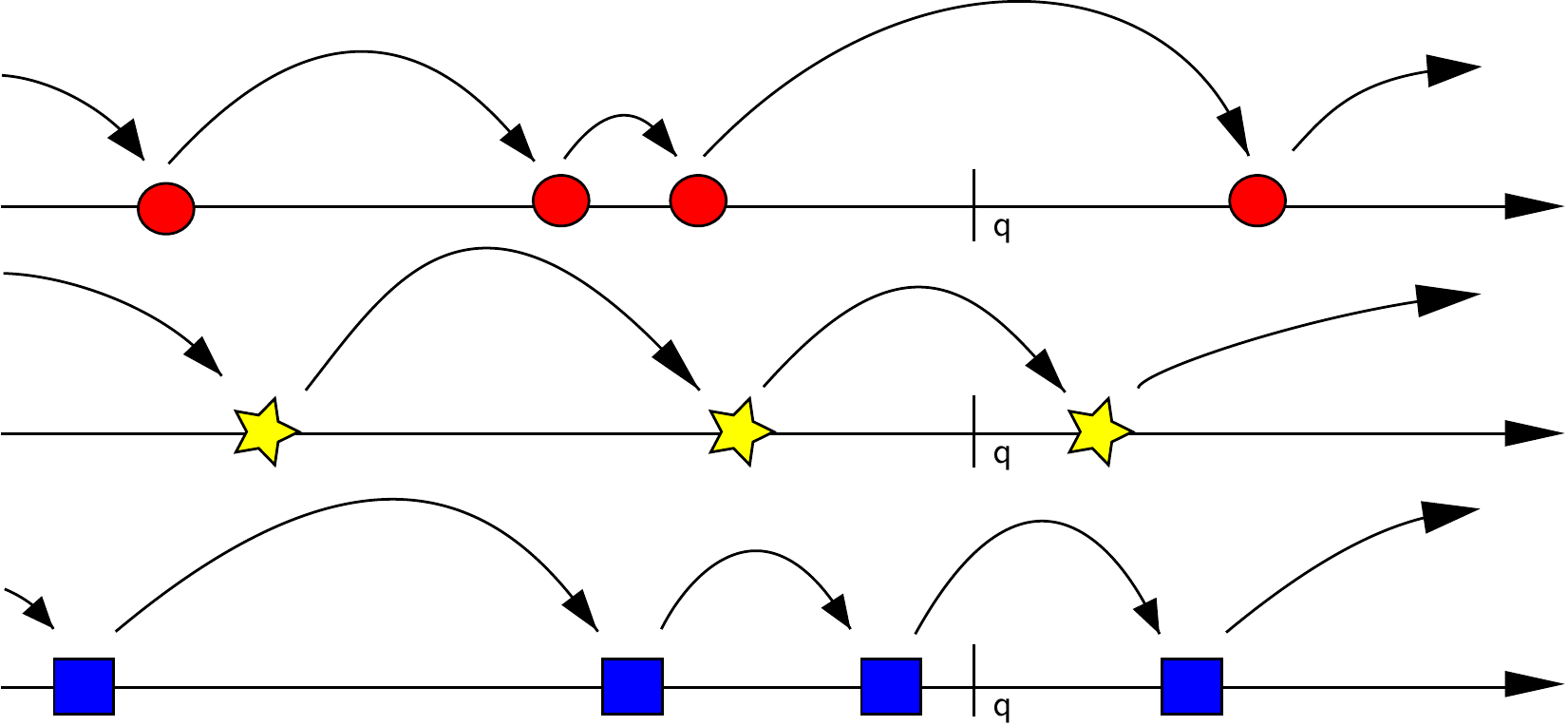}}\\
\subfloat[Marked Point Process realisation]{\includegraphics[width=\textwidth]{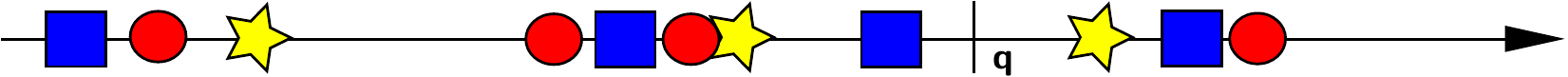}}
\caption{Sketch of algorithms \ref{algo_atteinte_defaillance} and \ref{algo_atteinte_defaillance_par}}
\end{minipage}
\hspace{1cm}
\hskip-20pt
\begin{minipage}[c]{0.45\textwidth}
\includegraphics[width=\textwidth]{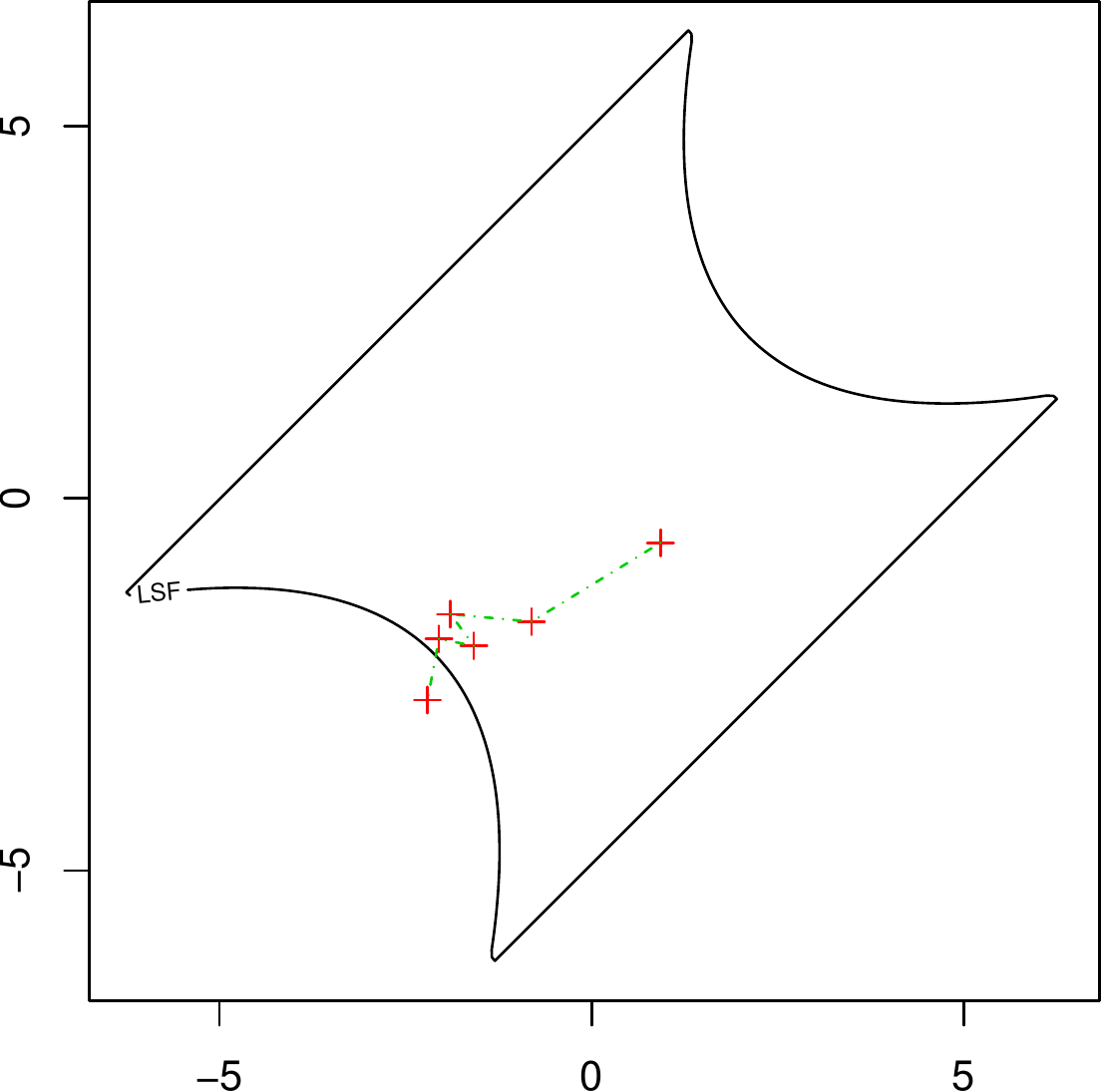}
\caption{Move of a particle}\end{minipage}
\end{figure}
\end{center}

\subsection{Practical implementation of algorithms \ref{algo_atteinte_defaillance} and \ref{algo_atteinte_defaillance_par}}
\label{subsection_practical_implementation_algo_par}

\paragraph{Simulating conditional distributions} We address here the issue of generating according to the conditional law $\mu^X( \, \cdot \mid g > q)$ for a given $q$. A general idea is to use convergence properties of an ergodic Markov Chain to its unique invariant probability to sample from a given distribution. Assuming $\mu^X$ has a \textit{pdf} $f_X$, it means we intend to generate a Markov Chain with stationary distribution $\one_{g > q} f_X$.

This implementation is especially simple when a reversible transition kernel $K$ is available. Guyader \textit{et al.} \cite{guyader2011simulation} presents two ways of getting it, referred as the Metropolis-Hastings method and a direct construction of the reversible kernel. We will then refer to the use of such a kernel starting from a given $\x$ as $K(\x, \cdot)$.
\begin{algo}[Metropolis-Hastings transition Kernel \cite{hastings1970monte}] \
\begin{itemize}
\item Generate $\mathbf{W}$ a standard multivariate Gaussian or symmetric Uniform distribution over a compact set of $\R^d$
\item Calculate $\mathbf{X}^* = \mathbf{x} + \sigma \mathbf{W}$
\item Calculate Metropolis-Hastings ratio: $\rho = \min\left(1, \dfrac{f_X(\X^*)}{f_X(\x)} \right) \one_{g(\X^*) > q}$
\item Accept the transition with probability $\rho$, return $\x$ otherwise
\end{itemize}
\end{algo}
\begin{algo}[Direct construction for standard Gaussian input space] \
\label{K_def}
\begin{itemize}
\item Generate $\mathbf{W}$ a standard multivariate Gaussian sample
\item Calculate $\mathbf{X}^* = \dfrac{\mathbf{x} + \sigma \mathbf{W}}{\sqrt{1 + \sigma^2}}$
\item Calculate $g(\X^*)$; return $\X^*$ if $g(\X^*) > q$, $\x$ otherwise
\end{itemize}
\end{algo}

\paragraph{The \textit{burn-in} and starting points parameters}
Because the goal is to reach the stationary state of the Markov Chain from a degenerated distribution, several transitions have to be done to insure independence between the starting point and the final sample and adequacy with the targeted distribution. This number of transitions is referred to as a \textit{burn-in} parameter and denoted by $T$. Eventually the last generated sample is kept. In theory one can start from any point provided the \textit{burn-in} is large enough but it is profitable to start with a point already following the targeted distribution as \textit{burn-in} will then serve only independence purpose.

\paragraph{Choice of the transition Kernel}
The examples presented Section \ref{examples} are all defined in the standard Gaussian space. In this context we will use the direct construction of the transition Kernel method (Algorithm \ref{K_def}) with $\sigma = 0.3$. The \textit{burn-in} parameter is set to 20 (usual value as in \cite{cerou2012sequential} or \cite{guyader2011simulation}).

\paragraph{Technical improvements \label{technical_improvements}}
Usual methods select a starting point randomly amongst points already following the targeted distribution. It appears that in practice it is particularly inappropriate to start from a point directly generated by the sample to move, called here a \textit{son}. In this context we adapt the random selection of the starting point for a moving particle to avoid all its sons and their possible replicas (if at one time the $T$ transitions were refused and so the new particle is simply a copy of the starting point). If at a given iteration there is no such starting point available, it starts from one of those and a move is counted only if at least one transition is accepted.

Furthermore if one has several cores available, it is possible to move simultaneously the $k$ last particles (sequential parallelisation of algorithm \ref{algo_atteinte_defaillance_par}). One recalls here that particles move independently and that moving a bunch of $N$ particles only allows for a better approximation of the stationary law of Markov Chains. Thus as soon as there are \emph{enough} particles \emph{above} the one to be moved, one can make the moves in parallel. Especially, $\chi^2$ tests were carried out in the examples of Section \ref{examples} on the numbers of iterations and they showed that the Poisson feature is very robust even for small $N$, \cad when the size of the population in which starting points is selected is small.

\paragraph{Use of a meta-model}
When the targeted distribution depends on the function of interest $g$, the Metropolis-Hastings sampling makes indeed the number of calls to $g$ $T$ times higher than what it \emph{should} be. This is to insure a good convergence of the Markov Chain to the stationary distribution. If one only wants to \emph{move} without any estimation of a statistical quantity, for instance to build a first DoE embedding failing samples for meta-model based algorithms, an approximation may be sufficient. In this scope we suggest the use of a cheap surrogate model to accept and/or reject transitions, true limit-state function being used only a limited number of times to \emph{control} the evolution of a particle. This will be illustrated Section \ref{section_DoE}.

\section{Application to extreme probabilities estimation}

\subsection{Description of the estimator} \label{pres_estim_proba}
In this case given $q$ we define $F = \{ \x \in \R^d \mid g(\x) > q\}$ and we intend to estimate $p = \proba{\X \in F}$. Let us first consider the ideal case: thanks to corollaries \ref{coro_nbr_mutations_1_particule} and \ref{coro_nbr_mutations_N_particules} we know that the sought probability is indeed directly related to the Poisson parameter of the number of moves to get realisations of $\X$ in $F$. Given $K$ independent realisations $(M_k)_k$ of a random variable following a Poisson law with unknown parameter $\lambda$, we thus propose to use the Maximum Likelihood Estimator of the parameter:
$$
\widehat{\lambda} = \dfrac{1}{K} \sum \limits_{k=1}^K M_k
$$
In our case we are not interested in $\lambda$ but $\exp (-\lambda/N)$, which leads to the following estimator for the probability:
$$
\widehat{p} = \exp \left( - \dfrac{\widehat{\lambda}}{N} \right) = \exp \left(-\dfrac{1}{KN} \right)^{\sum \limits_{k=1}^{K} M_k}
$$
%Since , one can calculate the moments of the estimator using the probability-generating function of a Poisson random variable:
%$$
%\begin{array}{rcl}
%\E{\widehat{p}} & = & \exp \left[K N \log(p) (1 - \exp(-1/KN) ) \right] \\[0.3cm]
%\cv{\widehat{p}}^2 & = &  \exp \left[- K N \log(p) (1 - \exp(-1/KN) )^2 \right] - 1
%\end{array}
%$$
Three remarks here:
\begin{itemize}
\item $\sum \limits_{k=1}^{K} M_k$ follows a Poisson law with parameter $-K N \log p$ and thus $K$ and $N$ are playing symmetric roles, which means that in the ideal case moving particles one by one or in cluster does not change anything.
\item Writing $x = KN$ we notice that making the asymptotic expansion of order 1 of $\exp(-1/x)$ gives the estimator proposed by Guyader \textit{et al.} \cite{guyader2011simulation}.
\item Furthermore considering $1-1/x$ instead of $\exp(-1/x)$ \textit{enforces} an unbiased estimator and gives a lower mean squared error $\E{(\p - p)^2} = \var{\p} + (\E{\p} - p)^2$.
\end{itemize}
Thus we chose the unbiased estimator for the probability $p$:
\begin{equation}
\widehat{p} = \left( 1 - \dfrac{1}{KN} \right)^{\sum \limits_{k=1}^{K} M_k}
\end{equation}
with $K$ the total number of algorithms run in parallel, $N$ the number of particles per algorithm, and $M_k$ the number of mutations for the k-th algorithm.

\subsection{Practical implementation} \label{impl_pratique_estim_proba}
In practice we can generate random variables $(M_k)_k$ of the number of mutations to get $N$ particles in $F$ and make the Poisson distribution approximation more accurate by increasing $N$ and $T$. Indeed there is a trade-off between the three parameters $K$, $N$ and $T$: the greater $K$ the faster the parallel estimation but the greater $N$ and $T$ the more robust.

Ideally, $N$ should be as small as possible to allow for a greater parallelisation but regarding the examples (cf Section \ref{examples}) $N$ should not be smaller than 10.

Furthermore, given $N_\text{call}$ the total number of calls to the limit-state function and $T$ the \textit{burn-in} parameter for conditional sampling, we have:
\begin{equation}
N_\text{call} = T  \{ \text{total number of iterations} \} \approx -T K N \log p = \alpha T \, \dfrac{-KN}{\alpha} \log p
\end{equation}
which means that for an equivalent computational budget, one can increase $T$ as much as one decreases $N$. Finally, given $n_c$ the number of cores available for calculation, the algorithm for estimating a probability writes as follows:
\begin{algo}[Algorithm for estimating a probability] \ \label{algo_estim_proba}
\begin{itemize}
\item Launch $n_c$ times in parallel algorithm \ref{algo_atteinte_defaillance_par} with stopping criteria $\min(\mathbf{q}_m) \geq q$ and $N \geq 10$
\item Count the total number of mutations $M = \sum \limits_{k=1}^{n_c} M_k$
\item Calculate $\widehat{p} = \left(1 - \dfrac{1}{N n_c} \right)^M$
\end{itemize}
\end{algo}

\subsection{Statistical analysis of the estimator \label{stat_estim_proba}}

From now on, let us consider that $N$ particles were moved, regardless of the number of algorithms practically used as this has no impact on ideal theoretical results. We then have the same results as \cite{guyader2011simulation} and reader is referred to this article for the proofs of the following results.

\paragraph{Moments of $\widehat{p}$ \label{moment_estim_proba}} One can calculate the moments of $\p$ using the probability-generating function of a Poisson random variable. We get:
\begin{equation}
\E{\widehat{p}} = p \, ; \,
\var{\widehat{p}} = p^2 \left( p^{\frac{-1}{N}} - 1 \right)
\end{equation}
This means this estimator almost achieves the Cramer-Rao bound: $-p^2 \log p /N$.
\paragraph{Confidence interval \label{intervalle_confiance_estim_proba}}
The discrete random variable $\p$ follows a Poisson distribution with parameter $-N\log p$. Furthermore we get $\log \p \loi \mathcal{P}(-N \log p)  \log (1-1/N)$. Considering $N \geq 10$ and $ p \leq 10^{-3}$ brings the following approximation:
\eq{
\log \p & \sim & \mathcal{N} \left( \log p, -\log p / N \right)
}

\begin{propo}
Given $\alpha \in [0,1]$ and $\za$ the quantile of order $1-\alpha/2$ of the standard normal distribution: $\proba{-\za < \mathcal{N}(0,1) < \za} = 1-\alpha$; an asymptotic confidence interval for $\p$ at $(1-\alpha)\%$ is given by:
\begin{equation}
\widehat{p} \exp \left(-\dfrac{\za^2}{2N} - \sqrt{\Delta} \right) \leq p \leq \widehat{p} \exp \left(-\dfrac{\za^2}{2N} + \sqrt{\Delta} \right)
\end{equation}
with:
$$
\Delta = \dfrac{\za^2}{N} \left( \widehat{t} + \dfrac{\za^2}{4N} \right) \text{ and } \widehat{t} = - \log \widehat{p}
$$
%
%\begin{proofprop}
%We get: $\widehat{t} = - \log \widehat{p} \sim \mathcal{N}(t, t/N)$ with $t =-\log p$. Let $\za$ be the $1-\alpha/2$ quantile of a standard Normal distribution, we have:
%$$
%\proba{\mid \widehat{t}-t \mid < \sqrt{t/N}  \za} = 1- \alpha = \proba{\widehat{t}^2 + t^2 - 2t \hat{t} < t/N \za^2}
%$$
%Solving this polynomial form on $t$ brings:
%$$
%\Delta = \left( \widehat{t}^2 + \dfrac{\za^2}{2N} \right)^2 - \widehat{t}^2 = \dfrac{\za^2}{N} \left( \widehat{t} + \dfrac{\za^2}{4N} \right) > 0
%$$
%Let $t_1$ and $t_2$ be the two roots, then:
%$$
%\proba{\mid \widehat{t}-t \mid < \sqrt{t/N}  \za} = \proba{t \in ]t_1 ; t_2 [} =  \proba{p \in ]p_1 ; p_2 [}
%$$
%with:
%$$
%\begin{array}{c}
%p_1 = - \log t_1 = \widehat{p} \exp \left(-\dfrac{\za^2}{2N} - \sqrt{\Delta} \right) \\[0.5cm]
%p_2 = -\log t_2 = \widehat{p} \exp \left(-\dfrac{\za^2}{2N} + \sqrt{\Delta} \right)
%\end{array}
%$$
%\end{proofprop}
\end{propo}

\subsection{Comparison with Monte-Carlo methods \label{comp_estim_proba}}

In this section we compare our algorithm with the ones presented in the introduction. We focus on the number of particles needed to gain a given precision on the estimator and on the computing time required to achieve it.

\paragraph{Coefficient of variation \label{calcul_nbr_appels_proba}}

The main difference between \cite{au2001estimation} and \cite{cerou2012sequential} stands in the sampling technique: while \cite{au2001estimation} uses Markov Chains, \cite{cerou2012sequential} uses particles methods developped by \cite{del2006sequential}. This leads \cite{au2001estimation} to introduce a coefficient $\gamma$ which characterises the correlation in between samples from the same Markov Chain (refer to the article for more details).

Considering $\delta$ the coefficient of variation (CV) of the estimator, we find in the corresponding articles in the first order in $N$:
\begin{itemize}
\item in \cite{au2001estimation}: $\delta_\text{Au}^2 = \dfrac{\log p}{\log p_0}  \dfrac{(1+\gamma)(1-p_0)}{N p_0}$
\item in \cite{cerou2012sequential}: $\delta_\text{Cer}^2 =\dfrac{1}{N} \left( \lfloor \dfrac{\log p}{\log p_0} \rfloor \dfrac{1-p_0}{p_0}+\dfrac{1-r_0}{r_0} \right)$
\item in \cite{guyader2011simulation}: $\delta_\text{Guy}^2 = \dfrac{-\log p}{N}$
\end{itemize}
For the purpose of our study we will consider the optimal setting $r_0 \approx p_0$ (like \cite{au2001estimation}) and drop the $\gamma$ coefficient to keep the following relation for Multilevel Splitting methods:
\eq{
\delta_\text{Au}^2 \approx \delta_\text{Cer}^2 \approx \delta_\text{Guy}^2 \approx \delta^2_\text{MS} \approx \dfrac{\log p}{\log p_0} \dfrac{1-p_0}{N p_0}
}
%Furthermore, \cite{au2001estimation} considers that $r_0 \approx p_0$ unlike \cite{cerou2012sequential}. Finally, neglecting $1$ in front of $-T \log p$ to get the order of magnitude we get for \cite{au2001estimation} and \cite{cerou2012sequential}
%\eq{
%\ncall \approx T N  \lfloor \dfrac{\log p}{\log p_0} \rfloor (1-p_0)
%}
%and so:
%\begin{itemize}
%\item in \cite{au2001estimation}: $\delta_\text{Au}^2 \approx\dfrac{T(\log p)^2}{\ncall} \dfrac{(1 + \gamma)(1-p_0)^2}{p_0 (\log p_0)^2}$
%\item in \cite{cerou2012sequential}: $\delta_\text{Cer}^2 \approx\dfrac{T(\log p)^2}{\ncall} \dfrac{(1-p_0)^2}{p_0 (\log p_0)^2}$
%\item in \cite{guyader2011simulation}: $\delta_\text{Guy}^2 \approx\dfrac{T(\log p)^2}{\ncall}$
%\end{itemize}
where $p_0 = 1 - 1/N$ in $\delta_\text{Guy}$.
\paragraph{Computational time}
Given a targeted precision $\delta$ on the estimator $\p$, we now focus on the computational time required to achieve it. Let $n_c$ be the number of available cores; one can consider that all operations are cost-free apart from calls to the \lsf and so computational time reduces to the number of calls to the \lsf done by each core.

\subparagraph{Naive Monte-Carlo algorithm}
In this algorithm there is no dependency between samples and so each core will have to do $\ncall/n_c$ calculations:
\eq{
t_{\text{MC}} & = & \lceil \dfrac{\ncall}{n_c} \rceil = \lceil \dfrac{1}{n_c \delta^2 p} \rceil
}

\subparagraph{Multilevel Splitting algorithms}
In Multilevel Splitting algorithms, there are $N$ samples generated initially and then $N(1-p_0)$ regenerated at each iteration. Thus, the computational time writes as follows:
\eq{
\tsmc & = & \dfrac{N}{n_c} + \lfloor \dfrac{\log p}{\log p_0} \rfloor T \left[ \dfrac{N(1-p_0)}{n_c} \wedge 1 \right] \\[0.4cm]
\tsmc & \approx & \dfrac{\log p}{\log p_0} \dfrac{1-p_0}{n_c \delta^2 p_0} + \dfrac{\log p}{\log p_0} T \left[ \dfrac{\log p}{\log p_0} \dfrac{(1-p_0)^2}{n_c \delta^2 p_0} \wedge 1 \right] \\
}
Depending on the parametrisation of the algorithm (choice of $N$ and $p_0$, number of cores $n_c$) we will have either $N(1-p_0) \geq n_c$ and so:
\eq{
\tsmc & \approx & \dfrac{T (\log p)^2}{n_c \delta^2} \dfrac{(1-p_0)^2}{p_0(\log p_0)^2 \label{tsmc_eq}}
}
or $N(1-p_0) \leq n_c$ and so:
\eq{
\tsmc & = & \dfrac{\log p}{\log p_0} T
\label{tsmc_non_par}}
Formula \eqref{tsmc_eq} is strictly decreasing in $p_0$ while \eqref{tsmc_non_par} is strictly increasing. This can eventually suggest an optimal value $p_0^*$ for $p_0$ depending on the sought probability $p$, the number of cores $n_c$ and the precision $\delta$, it is the solution of:
\eq{
\dfrac{\log p}{\log p_0^*} \dfrac{(1-p_0^*)^2}{n_c \delta^2 p_0^*} = \dfrac{N(1-p_0^*)}{n_c} = 1
}
Finally we can write:
\eq{
\tsmc(p_0) \geq \tsmc(p_0^*) = \dfrac{T (\log p)^2}{n_c \delta^2} \dfrac{(1-p_0^*)^2}{p_0^*(\log p_0^*)^2} > \dfrac{T (\log p)^2}{n_c \delta^2}
}
%Sequential Monte-Carlo algorithms are known to allow for sequential parallelisation, which means that at each threshold the $N(1-p_0)$ samples can be generated in a parallel way. For \cite{au2001estimation} and \cite{cerou2012sequential} on can consider that $n_c < N(1-p_0)$ and thus:
%\eq{
%t_\text{Au} & = & \lceil \dfrac{\ncall}{n_c} \rceil \approx\dfrac{T(\log p)^2}{n_c \delta^2} \dfrac{(1 + \gamma)(1-p_0)^2}{p_0 (\log p_0)^2} \\
%t_\text{Cer} & \approx & \dfrac{T(\log p)^2}{n_c \delta^2} \dfrac{(1-p_0)^2}{p_0 (\log p_0)^2}
%}
%
%Since  Guyader \textit{et al.} algorithm is a limit case of sequential Monte-Carlo with $p_0 = 1-1/N$, $N(1-p_0) = 1$, no parallelisation is possible and it remains:
%\eq{
%t_\text{Guy} & \approx & \lceil \dfrac{T(\log p)^2}{ \delta^2 } \rceil
%}

\subparagraph{Moving Particles algorithm}
There we have $N/n_c$ particles per algorithm \ref{algo_atteinte_defaillance_par} and so the "arrival times" of the $n_c$ algorithms will be distributed as the realisations of $n_c$ Poisson laws with parameter $-N/n_c \log p$.

\begin{propo} \label{tpar_proba}
Let $t_\text{par}$ be the random variable of the effective computing time for full parallel algorithm \ref{algo_estim_proba}, we have:
\eq{ \label{comput_time_proba}
\E{t_\text{par}} & = & \dfrac{T (\log p)^2}{n_c \delta^2} \left(1 + \sqrt{\dfrac{n_c \delta^2}{(\log p)^2}}\sqrt{2 \log n_c} + \dfrac{1}{T \log 1/p} \right) \\
}

\begin{proofprop}
Let $\lambda$ be the parameter of the Poisson laws: $\lambda = -N/n_c \log p$. In the extreme value theory framework, we are interested in the so called location parameter $b_n$ which drives the mean of the maximum of $n$ iid random variables with \textit{cdf} $F$. It is the solution of the equation:
$$
b_n = F^{-1}\left( 1 - \dfrac{1}{n} \right)
$$
For the Normal distribution, we have:
$$
b_n = \sqrt{2 \log n} + \dfrac{\log \log n + \log 4 \pi}{2 \sqrt{2 \log n}} \sim \sqrt{2 \log n}
$$
Furthermore, we know that Normal approximation of a Poisson distribution with parameter $\lambda$ is valid in the range $(-\sqrt{\lambda}, \sqrt{\lambda})$. Here, this means that as soon as $\sqrt{2 \log n_c} < \sqrt{\lambda}$, we can use the approximation of the Poisson law by a Normal one's to calculate the constant. The practical values of $N/n_c \approx 10^1$, $- \log p \approx 10^1$ and $n \approx 10^2$ allows us to make the approximation:
$$
\mathcal{P}(\lambda) \sim \mathcal{N}\left(\lambda,\lambda \right)
$$
Let $N_\text{max}$ be the random variable of the maximum of $n_c$ iid standard Gaussian variables. The total number of calls is the sum of the $N/n_c$ initial calls to the limit-state function and the number of iterations. Thus we have:
$$
\E{t_\text{par}} = \E{ T  ( N_\mx  \sqrt{\lambda} + \lambda) + N/n_c} \approx T \left( \dfrac{N}{n_c} \log 1/p + \sqrt{\dfrac{N}{n_c} \log 1/p} \sqrt{2 \log n_c} \right) + \dfrac{N}{n_c}
$$
and so in terms of coefficient of variation:
$$
\E{t_\text{par}} = \dfrac{T}{n_c \delta^2} (\log p)^2 + T \sqrt{\dfrac{(\log p)^2}{n_c \delta^2}} \sqrt{ 2 \log n_c} + \dfrac{-\log p}{n_c \delta^2}
$$
\end{proofprop}
\end{propo}
The additional term due to the full parallelisation would drop in case of sequential parallelisation. Indeed considering the "arrival times" are normally distributed, there will be as many \emph{shorter} than \emph{longer} algorithms comparing to the reference value $-N/n_c \log p$. If one can afford sequential parallelisation (cf. Section \ref{technical_improvements}), shorter and longer algorithm will compensate each other. So for comparison with classical Sequential Monte-Carlo one can retain:
\eq{
\tpar = \dfrac{T (\log p)^2}{n_c \delta^2}
}

\paragraph{Conclusion on the estimator}
On the one hand we have found an optimal value of $p_0$ for classical Multilevel Splitting algorithm in terms of computational time (considering that the only important operation is a call to the limit-state function). On the other hand we see that our approach gives always a better result than the Multilevel Splitting method with the optimal $p_0$. In other words, our approach allows for taking $p_0 \rightarrow 1$ while keeping the parallel computation. Thus this is the optimal way of computing Multilevel Splitting methods.

Furthermore, with standard values of $\log 1/p \approx 10^1$, $n_c \approx 10^2$ and $\delta^2 = 10^{-2}$, we get $(\log p)^2 / (n_c \delta^2) \approx 10^2$, which means $\tpar$ is multiplied by $\approx 1.1$ if no sequential parallelisation is applied. As sequential parallelisation can indeed increase computational time consequently, this result is of great interest because it shows that without losing too much time the implementation of Multilevel Splitting methods can be a lot easier.

\begin{figure}
\centering
\includegraphics[scale=0.3]{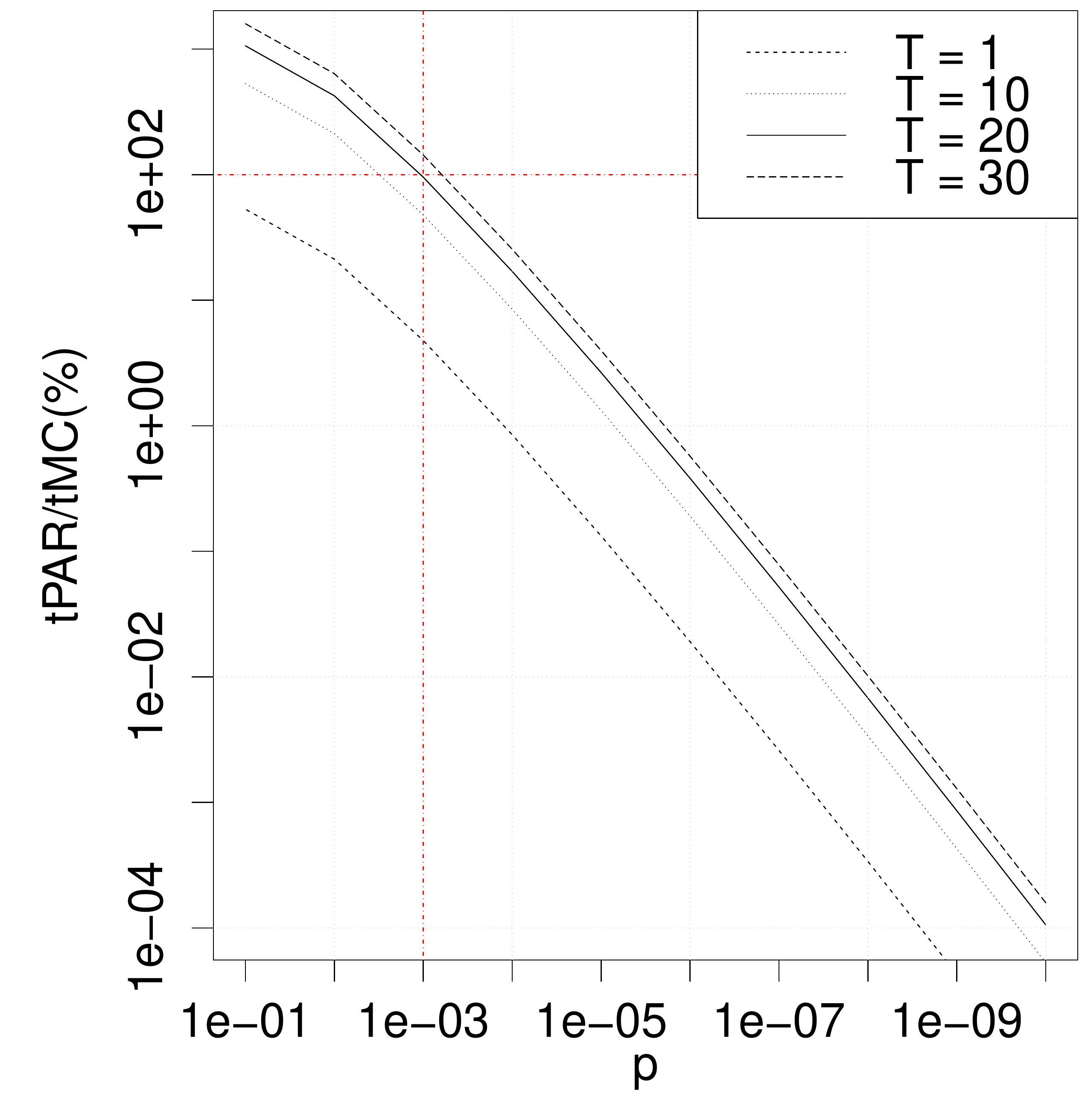}
\caption{$\tpar/\tmc(\%)$ against $p$ for different \textit{burn-in} parameters}
\label{tPAR_tMC}
\end{figure}

Finally we compare $\tmc$ and $\tpar$:
\eq{
%\log \left( \dfrac{\tpar}{\tmc} \right) = \log \left(T p (\log p)^2 \left( 1 + \sqrt{\dfrac{n_c \delta^2}{(\log p)^2}}\sqrt{2 \log n_c} + \dfrac{1}{T \log 1/p} \right) \underset{p \rightarrow 0}{\sim} & \log(p)
\log \left( \dfrac{\tpar}{\tmc} \right) \underset{p \rightarrow 0}{\sim} \log \left(T p (\log p)^2 \right)
}
Figure \ref{tPAR_tMC} plots $t_\text{par} / t_\text{MC} (\%)$ for several values of $T$ in logarithmic scale. It is clearly visible that the relation becomes linear when $p$ becomes small and especially for a standard value of $T = 20$, our algorithm makes better than a naive Monte-Carlo as soon as $p \lesssim 10^{-3}$.

\section{Application to extreme quantile estimation}

\subsection{Description of the estimator}

In this case we define $F = \{ \x \in \R^d \mid g(\x) > q \}$ and $p = \proba{\X \in F}$ and we intend to estimate $q$.
Let us first suppose we can have the infinite sequence of the events $(T_m)_m$ of the Poisson Process (with $T_0 = 0$) and derive some results. We focus on the counting random variable $M_t = \sup \{ m \geq 0 \mid T_m \leq t \}$ at time $t = - \log p$. We know that $M_t \loi \mathcal{P}(-N \log p)$.

\begin{lemme}[Laws of random variables $T_{M_t}$ and $T_{M_t + 1}$] \label{loi_fmt}
Let us denote by $F_{M_t}$ and $F_{M_t +1}$ the \textit{cdf} associated to $T_{M_t}$ and $T_{M_t +1}$ respectively. The following result holds:
\eq{
\forall (\alpha, \beta) \in \R_+^2, \proba{ (\tmtun - t > \alpha/N) \cap (t - \tmt \geq \beta/N)} = e^{-\alpha}  e^{-\beta} \, \one_{[0; Nt)}(\beta)
}

\begin{prooflemme}
Given $(\alpha, \beta) \in \R_+^2$, we write $\forall k \in \N, \Delta_k = \{ (T_{k+1} - t > \alpha/N) \cap (t - T_k \geq \beta/N) \}$. We have:
\begin{align*}
\proba{\Delta_{M_t} } &= \sum \limits_{k=0}^\infty \proba{ \Delta_{M_t} \cap \{ M_t = k \} }
\end{align*}
Noticing here that $\{M_t = k \} = \{T_k \leq t < T_{k+1} \}$ we have:
$$
\proba{\Delta_{M_t} } = \sum \limits_{k=0}^\infty \proba{ \Delta_k } \\
$$
With $\forall k \in \N^*, f(t_k, t_{k+1}) = N^{k+1} e^{-N t_{k+1}} t_k^{k-1}/(k-1)! \one_{0<t_k<t_{k+1}}$ we get:
$$
\forall k \in \N, \proba{\Delta_k } = \dfrac{(N(t-\beta/N))^k}{k!} e^{-Nt-\alpha}  \one_{[0; t)}(\beta/N) \one_{[0; +\infty)} (\alpha) 
$$
and so the result announced:
$$
\proba{(\tmtun - t > \alpha/N) \cap (t - \tmt \geq \beta/N) } = e^{-\alpha} \one_{[0; +\infty)} (\alpha)  e^{-\beta} \one_{[0; Nt)}(\beta)
$$

\end{prooflemme}
\end{lemme}

\begin{coro} \label{pos_centre_t}
The center of the interval $[\tmt; \tmtun]$ converges toward a random variable centred in $t$ with symmetric \textit{pdf}, \cad:
\eq{
N \left( \dfrac{\tmtun + \tmt}{2} - t \right) \tend{Nt} Z
}
with $Z$ a random variable with \textit{pdf} $f_Z(z) = e^{- 2 \mid z \mid}$

\begin{proofcoro}
From the joint probability distribution of $\tmt$ and $\tmtun$ we have:
$$
\forall (\alpha, \beta) \in \R_+^2, \proba{ (\tmtun - t > \alpha/N) \cap (t - \tmt \geq \beta/N)} \tend{Nt} e^{-\alpha} \, e^{-\beta}
$$
\cad:
$$
N(\tmtun - t ; t - \tmt ) \conv{Nt} (Z_1, Z_2)
$$
with $Z_1$ and $Z_2$ iid with distribution $\mathcal{E}(1)$. The difference between the center of the interval and $t$ is thus the difference between two independent random variables iid with distribution $\mathcal{E}(2)$, which gives the desired result.

\end{proofcoro}
\end{coro}
From Lemma \ref{loi_fmt} and Corollary \ref{pos_centre_t} we can define the following quantity of interest:
$$
\widetilde{q} = \dfrac{1}{2} q_{M_t} + \dfrac{1}{2} q_{M_t + 1}
$$
with $(q_m)_m = (\Lambda^{-1}(T_m))_m$. We assume that the \textit{pdf} $f$ of $g(\X)$ is continuous at $q$; knowing the law of $(\tmt, \tmtun)$ and making an asymptotic expansion around $t$ we can calculate the moments of $\widetilde{q}$ and get:
\begin{align}
\E{\widetilde{q}} &= q + o\left(\dfrac{1}{N} \right) \\
\var{\widetilde{q}} &= \dfrac{p^2}{f(q)^2}\dfrac{1}{N^2} + o\left(\dfrac{1}{N^2} \right)
\end{align}
\subsection{Practical implementation}

Unfortunately we do not observe the $(T_m)_m$ but only $q_m = \Lambda^{-1}(T_m)$ for which we do not have any expression and we cannot observe $M_t$ nor $T_{M_t}$ and $T_{M_t + 1}$. However we know that $M_t \loi \mathcal{P}(-N \log p)$; writing $M = \lceil -N \log p \rceil$, we then choose as an estimator for $q$:
\begin{equation}
\widehat{q} = \dfrac{1}{2} q_{M-1} + \dfrac{1}{2} q_{M}
\end{equation}
Indeed we want to estimate $q$ by looking at some particular events defined by their rank. Especially if several algorithms are run in parallel, they will be events of the full Process: $n_c$ computers available brings $M = \lceil N n_c \log p \rceil$. However to rebuild the full process until event number $M$ requires to make sure that all realisations $(T_m)_m$ have overpassed a given time $t$ and that the total number of events before that time $t$ will be greater than $M$. We propose here two possible ways to achieve this with parallel algorithms: a 2-passes algorithm and a sequential one. The sequential algorithm is optimal considering the number of calls to the \lsf but allows only for a sequential parallelisation, which can indeed turn it into a longer one's. Denoting by $(q_m^i)_m$ the sequence of the successive minima of algorithm number $i \in \llbracket 1; n_c \rrbracket$, the algorithms are:
\begin{algo}[A 2-passes algorithm for quantile estimation] \ \label{algo_estim_quantile_2fold}
\begin{itemize}
\item Do $n_c$ times algorithm \ref{algo_atteinte_defaillance_par} with stopping criteria "number of moves = $m_0$" and $N \geq 10$
\item Get the maximal value reached: $q_\text{max} = \underset{i = 1..n_c}{\max} q_{m_0}^i$
\item Restart the $n_c$ algorithms with stopping criteria "$\min(\mathbf{q}_m) \geq q_\text{max}$"
\item Merge the $n_c$ sequences $(q^i_m)_m$ and sort the resulting one; call it $(q_m)_m$
\item Calculate $\qcl = \dfrac{1}{2} \left( q_{M-1} + q_M \right)$
\end{itemize}
\end{algo}
The choice of $m_0$ will be discussed further section \ref{comp_quantile_MC}. One can already notice that depending on this value it will be certain ($m_0 = \lceil -N \log p \rceil$) or not to eventually get enough events.
\begin{algo}[A sequential algorithm for quantile estimation] \ \label{algo_estim_quantile_seq}
\begin{itemize}
\item Do $n_c$ times algorithm \ref{algo_atteinte_defaillance_par} with stopping criteria "number of moves = 1" and $N \geq 10$
\item Get the minimal value reached: $q_\text{min} = \underset{i = 1..n_c}{\min} q_1^i$
\item $k = 1$. While $\# \{ q_j^i \mid q_j^i \leq q_\text{min} ; i = 1..n_c,  j = 1..k \} < M$
\begin{itemize}
\item[+] $k := k + 1$
\item[+] Restart the $n_c$ algorithms with stopping criteria "number of moves = 1"
\item[+] Get the minimal value reached: $q_\text{min} = \underset{i = 1..n_c}{\min} q_k^i$
\end{itemize}
\item Merge the $n_c$ sequences $(q^i_m)_m$ and sort the resulting one; call it $(q_m)_m$
\item Calculate $\qcl = \dfrac{1}{2} \left( q_{M-1} + q_M \right)$\end{itemize}
\end{algo}
The sequential version of the algorithm insures a minimum number of iterations as it stops as soon as there are enough events. However it needs communication between the $n_c$ algorithms which can finally slow down the whole estimation.

\subsection{Statistical analysis of the estimator} \label{stat_estim_quantile}

\paragraph{Some general results}
Let $N$ be the total number of particles and $m = \lceil - N \log p \rceil$. We first present some asymptotic results for an estimator $q_{m+k} = \Lambda^{-1}(T_{m+k})$ with a given $k$ as $N \rightarrow +\infty$. Then we study the property of a linear combination of $\qmk$ around $m$. From now on, let us suppose that $g(\X)$ has \textit{cdf} $F$ and \textit{pdf} $f$ continuous at $q$.

\begin{propo}[Central Limit Theorem]
If $f(q) \neq 0$, then:
\eq{
\sqrt{N} \left( q_{m+k} - q \right) \tend{N} \mathcal{N}\left( 0, \dfrac{-p^2 \log p}{f(q)^2} \right)
}

\begin{proofprop}
This proof comes mainly from \cite{guyader2011simulation}.

We have $q_{m+k} = \Lambda^{-1}(T_{m+k})$ where $\tmk \sim \Gamma_{m+k} / N$ with $\Gamma_{m+k}$ a Gamma random distribution with parameter $m+k$. From the definition of $m = \lceil -N \log p\rceil$ we get:
$$
\dfrac{m+k}{N} \underset{N \rightarrow + \infty}{\longrightarrow} -\log p
$$
which lets us rewrite the Central Limit Theorem for the Gamma random variable as follows:
$$\sqrt{N} \, \dfrac{\tmk - (-\log p)}{\sqrt{-\log p}} \tend{m} \mathcal{N}(0,1)$$
One eventually concludes by making an asymptotic expansion of $\Lambda(\qmk)$ around $q$:
$$
\Lambda(\qmk) - \Lambda(q) = (\qmk - q)  \Lambda'(q) + o(\qmk - q) = (\qmk - q)  f(q)/p + o(\qmk - q)
$$
then:
$$
\sqrt{N} (\qmk -q) \tend{m} \mathcal{N} \left( 0,\dfrac{-p^2 \log p}{f(q)^2} \right) \\[0.5cm]
$$
\end{proofprop}
\end{propo}

\begin{propo}[Bounds on bias]
With the same hypothesis as for a naive Monte-Carlo estimator (see for example \cite{arnold1992first} p.128) and $f'(q) < 0$ we get the following boundaries for the bias:
\eq{
\E{\qmk} - q & \geq & \dfrac{p}{Nf(q)} \left( \dfrac{-\log p}{2} \left( 1 + \dfrac{f'(q) p}{f(q)^2} \right) + k\right) + o(\dfrac{1}{N}) \\[0.5cm]
\E{\qmk} - q & \leq & \dfrac{p}{Nf(q)} \left( \dfrac{-\log p}{2} \left( 1 + \dfrac{f'(q) p}{f(q)^2} \right) + k+ 1 \right) + o(\dfrac{1}{N})
}

\begin{proofprop}
Let us write the asymptotic expansion of $\qmk$ around $\tmk$:
\begin{align*}
\qmk &= \Lambda^{-1}(\tmk) \\
&= \Lambda^{-1}(t) + (\tmk - t)(\Lambda^{-1})'(t) + \dfrac{(\tmk - t)^2}{2}(\Lambda^{-1})''(t) + o_\mathbb{P} \left( (\tmk - t)^2 \right) \\
\qmk - q &= \dfrac{p}{f(q)} (\tmk - t) + \dfrac{(\tmk - t)^2}{2} \left( -\dfrac{p}{f(q)} \right) \left( 1 + \dfrac{p f'(q)}{f(q)^2} \right) + o_\mathbb{P} \left( (\tmk - t)^2 \right)
\end{align*}
We have:
$$
\E{\tmk - t} = \dfrac{m+k}{N} - t = \dfrac{1}{N}(m - Nt + k) \in [\dfrac{k}{N}; \dfrac{k+1}{N} ) \\
$$
and:
\begin{align*}
\E{(\tmk - t)^2} &= \var{\tmk} + (\dfrac{m+k}{N}-t)^2 = \dfrac{m}{N^2} + \dfrac{k}{N^2} + \dfrac{1}{N^2}(m -Nt + k)^2 \\
\E{(\tmk - t)^2} & \in \dfrac{t}{N} + \dfrac{1}{N^2} [ k + k^2, k+1 + (k+1)^2)
\end{align*}
which concludes the proof.
%$$
%\begin{array}{rcl}
%\qmk & = & S^{-1}(\exp-\tmk) \\[0.4cm]
% & = & S^{-1}(\exp -t) + (\exp (-\tmk) - \exp(-t)) (S^{-1})'(\exp-t) \\[0.4cm]
% & & + \, \dfrac{(\exp (-\tmk) - \exp(-t))^2}{2} (S^{-1})''(\exp-t) + o_\mathbb{P} \left( (\exp (-\tmk) - \exp(-t))^2 \right) \\[0.4cm]
%\qmk & = & q + \dfrac{-\left( \exp (-\tmk) - p \right)}{f(q)} + \dfrac{(\exp (-\tmk) - p)^2}{2} \dfrac{-f'(q)}{f(q)^3} + o_\mathbb{P} \left( (\exp (-\tmk) - p)^2 \right) \\[0.4cm]
%\end{array}
%$$
%We have:
%$$\E{\exp (-\tmk)^i} = \exp \left( -(m+k) \log (1+i/N) \right)
%$$
%Then, using again the following inequality:
%$$-N \log p \leq m \leq -N \log p +1$$
%we get the bounds on the expectations and eventually the result.
\end{proofprop}
\end{propo}

\begin{propo}[Confidence interval] \label{Intervalle_confiance_quantile}
Writing $\za$ the $(1-\alpha)\%$ quantile of a standard Gaussian distribution, $m_- = \lfloor m - \za \sqrt{m} \rfloor$ and $m_+ = \lceil m + \za \sqrt{m} \rceil$, we have:
\begin{equation}
\proba{q \in [ q_{m_-}, q_{m_+} ] } \conv{m} 1 - \alpha
\end{equation}

\begin{proofprop}
Considering the approximation of a Poisson distribution by a Normal distribution: $M_t \sim \mathcal{N}(m,m)$
and writing $\za$ the $1-\alpha$ \% quantile of a standard Gaussian law:
$$
\proba{M_t \in [ m - \za \sqrt{m}, m + \za \sqrt{m}]} = 1- \alpha
$$
We conclude by noticing that the sequence of $(q_i)_i$ is strictly increasing.
\end{proofprop}

\end{propo}

\begin{propo}[Multidimensional Central Limit Theorem] \label{TCL_vectoriel}
Let us now consider the vector $(q_m, q_{m+1}, \dots, q_{m+k})$. We can write the following multidimensional central limit theorem:
\begin{equation}
\sqrt{N} \left[ \left( \begin{array}{c} q_{m} \\ \vdots \\ q_{m+k} \end{array} \right) - q \left( \begin{array}{c} 1 \\ \vdots \\ 1 \end{array} \right) \right] \tend{N} \mathcal{N} \left( \left( \begin{array}{c} 0 \\ \vdots \\ 0 \end{array} \right), \dfrac{-p^2 \log p}{f(q)^2} \left( \begin{array}{ccc} 1 & \dots & 1 \\ \vdots & \ddots & \vdots \\ 1 & \dots & 1 \end{array} \right) \right)
\end{equation}

\begin{proofprop}
The $(T_{m+k})_k$ are the times of the Poisson Process with parameter $N$ such that we know their joint probability distribution and get:
$$ \forall k \in \mathbb{Z} \mid m+k > 0, \cov{T_{m}}{T_{m+k}} = \dfrac{m}{N^2}$$

Let us now define $\phi : x \mapsto = \Lambda^{-1}(x)$ and calculate the covariance matrix between the $(q_{m+k})_k$. Since $\qmk = q + (T_{m+k}-t) \phi'(t) + o(T_{m+k} - t)$ and $\phi'(t) = p/f(q)$, we find:
$$
\cov{q_m}{q_{m+k}} = \left( \dfrac{p}{f(q)} \right)^2 \dfrac{m}{N^2} + o(\dfrac{1}{N})
$$
from which it comes:
$$ N \cov{q_m}{q_{m+k}} \underset{N \rightarrow \infty}{\longrightarrow} \dfrac{-p^2 \log p}{f(q)^2}$$
which concludes the proof.
\end{proofprop}
\end{propo}

\paragraph{Statistical properties of the quantile estimator}
\begin{propo}\label{stat_estim_q}
The estimator $\qcl = \dfrac{1}{2}(q_{m-1} + q_m)$ has the following properties:
\begin{equation}
\sqrt{N} \left( \qcl - q \right) \tend{N} \mathcal{N}\left( 0, \dfrac{-p^2 \log p}{f(q)^2} \right)
\end{equation}
and:
\begin{equation} \label{quantile_bound_bias}
\begin{array}{rcl}
\E{\qcl} - q & \geq & \dfrac{p}{Nf(q)} \left( \dfrac{-\log p}{2} \left( 1 + \dfrac{f'(q) p}{f(q)^2} \right) - \dfrac{1}{2} \right) + o(\dfrac{1}{N}) \\[0.5cm]
\E{\qcl} - q & \leq & \dfrac{p}{Nf(q)} \left( \dfrac{-\log p}{2} \left( 1 + \dfrac{f'(q) p}{f(q)^2} \right) + \dfrac{1}{2} \right)+ o(\dfrac{1}{N}) 
\end{array}
\end{equation}
\end{propo}

\begin{rem}
For exponential tails, \cad distributions with \textit{pdf} $f(x) \underset{+\infty}{\sim} \exp ( - ( \frac{\mid x \mid}{x_0} )^k )$ for some $k \in \N$ and $x_0 \in \R^*_+$ one have:
$$
\dfrac{-\log p}{2} \left( 1 + \dfrac{f'(q) p}{f(q)^2} \right) \underset{p \rightarrow 0}{\sim} \dfrac{-1}{2} \dfrac{k-1}{k}
$$
with equality if $k = 1$. This means that for exponential tails, the bounds on bias simplify as follow:
\begin{equation} 
-\dfrac{1}{2}\dfrac{p}{f(q)} \left( 2 - \dfrac{1}{k} \right) \leq \underset{N}{\liminf} \, N (\E{\qcl} - q) \leq  \underset{N}{\limsup} \, N (\E{\qcl} - q) \leq \dfrac{1}{2k}\dfrac{p}{f(q)}
\end{equation}
\end{rem}

\subsection{Comparison with Monte-Carlo methods} \label{comp_quantile_MC}
As for the probability estimator, we now want to benchmark the computing time of our algorithm given a precision. Unlike in Section \ref{comp_estim_proba} we can only compare with naive Monte-Carlo as Multilevel Splitting methods are not able to produce quantile estimators and Guyader \textit{et al.} algorithm does not allow for parallelisation.

\paragraph{Naive Monte-Carlo algorithm}
Naive Monte-Carlo estimator has the following properties:
\begin{equation}
\sqrt{N} \left( \widehat{q}_\text{MC} - q \right) \tend{N} \mathcal{N}\left(0, \dfrac{p(1-p)}{f(q)^2} \right)
\end{equation}
with $N$ the number of samples and $f$ the \textit{pdf} of $g(\X)$.

This brings for the coefficient of variation:
\begin{equation}
\delta^2 = \dfrac{p(1-p)}{N f(q)^2} \dfrac{1}{q^2} + o(\dfrac{1}{N}) \approx \dfrac{p}{N q^2 f(q)^2} 
\end{equation}
and then the computing time:
\eq{
t_\text{MC} & = & \lceil \dfrac{N}{n_c} \rceil = \lceil \dfrac{p}{q^2 f(q)^2 \delta^2 n_c} \rceil
}

\paragraph{Moving Particles algorithm}
Both versions of the algorithm (\ref{algo_estim_quantile_2fold} and \ref{algo_estim_quantile_seq}) produce finally the same estimator and so have the same statistical properties:
\eq{
\delta^2 = \dfrac{-p^2 \log p}{n_c N q^2 f(q)^2} + o\left(\dfrac{1}{n_c N}\right)
}
with $N$ the number of particles per algorithm and $n_c$ the number of cores.

The computing time of each algorithm depends on the number of iterations done by each parallel algorithm; especially for algorithm \ref{algo_estim_quantile_2fold} this number will be driven by the parameter $m_0$.

\subparagraph{Number of iterations of Algorithm \ref{algo_estim_quantile_2fold}}
After the first pass we have indeed $n_c$ iid random variables with distribution $\Gamma_{m_0}/N$ random variables for which we consider the maximum $T_\mx$. In the second pass, we move particles until time $T_\mx$ and so the total number of iterations follows a Poisson law with parameter $N T_\mx$.

\begin{propo}
Let $N_\text{iter}$ be the random variable counting the number of iterations of algorithm \ref{algo_estim_quantile_2fold}, we have:
\eq{
\E{N_\text{iter}} = m_0 + \sqrt{2 m_0 \log n_c }
}

\begin{proofprop}
We apply here the same line of argumentation as in the proof of Proposition \ref{tpar_proba}, which let us approximate the Gamma distribution by a Gaussian one:
$$
\dfrac{\Gamma_{m_0}}{N} \sim \mathcal{N}\left( \dfrac{m_0}{N}, \dfrac{m_0}{N^2} \right)
$$
Let $N_\text{max}$ be the random variable of the maximum of $n_c$ standard Gaussian variables, we have:
$$
\E{ T_\text{max} } = \E{ N_\mx  \dfrac{\sqrt{m_0}}{N} + \dfrac{m_0}{N} } \approx \dfrac{m_0}{N} + \dfrac{\sqrt{m_0}}{N} \sqrt{2 \log n_c}
$$
then
$$
\E{\Niter} = \E{ \E{\Niter \mid T_\mx} } = N \E{T_\mx} = m_0 + \sqrt{2 m_0 \log n_c}
$$

\end{proofprop}
\end{propo}

We now detail a criteria for choosing $m_0$. Indeed, we consider that one can accept to take a risk $\alpha$ that the Poisson Process does not go further enough. Let $M_t$ be the counting variable of the marked Poisson Process a time $t$ and $m = \lceil -n_c N \log p \rceil$ the targeted number of events, the criteria writes as follows:
\eq{
\proba{ M_{T_\mx} \leq m } \leq \alpha
}
\begin{propo}[Choice of $m_0$ \label{choix_m0}]
With the previous notations and $t = -\log p$, we have:
\eq{
m_0 = \lceil Nt + \beta^2 / 2 - \beta \sqrt{\Delta} / 2 \rceil
}
with:
$$
\beta = b_{n_c} - \dfrac{\log \log 1/\alpha}{\sqrt{2 \log n_c}}
$$
$b_{n_c}$ being the localisation parameter of the Gaussian law in the framework of Extreme Value Theory: $b_{n_c} = \sqrt{2 \log n_c} - \left( \log \log n_c + \log 4 \pi \right) / 2\sqrt{2 \log n_c}$ and
$$
\Delta = \beta^2 + 4Nt
$$

\begin{proofprop}
We have:
$$
\proba{ M_{T_\mx} \leq m }= \proba{ T_\mx \leq T_m }
$$
Furthermore, $T_m = \Gamma_m / n_c N \sim \mathcal{N}(t, t/ n_c N)$. So
$$
\proba{ T_\mx \leq T_m } \approx \proba{ T_\mx \leq t}
$$
Finally, we seek for $m_0$ such as:
$$
\proba{ T_\mx \leq t } \leq \alpha
$$
Approximating once again Gamma laws with Gaussian distribution and using the extreme value theorem, we get:
$$
\alpha = \exp \left( - \exp \left( -\sqrt{2 \log n_c} \left( \dfrac{Nt}{\sqrt{m_0}} - \sqrt{m_0} - b_{n_c} \right) \right) \right)
$$
which concludes the proof.
\end{proofprop}
\end{propo}

\begin{rem}The targeted value $\alpha$ should not be set too small as the approximation of Gamma laws with Gaussian distributions is not correct for rare events. However we know that taking $m_0 = \lceil -N \log p \rceil$ ensures a sufficient number of iterations because of $n_c \lceil - N \log p \rceil \geq \lceil - n_c N \log p \rceil$. Furthermore, if after Algorithm \ref{algo_estim_quantile_2fold} the number of events is not sufficient, one can restart it for the number of missing events.
\end{rem}

\begin{propo}[Expectation of $\Niter$ in the 2-passes algorithm\label{esp_niter_quantile}]
With this value of $m_0$, we have:
\eq{
\E{\Niter} \approx - N \log p + \sqrt{2 N \log 1/p \log n_c}
}
\end{propo}

\subparagraph{Number of iterations of Algorithm \ref{algo_estim_quantile_seq}}
In this sequential algorithm there is no parameter $m_0$ to set. Writing $T_\mn(k)$ the minimum of $k$ iid random variables with distribution $\Gamma_k / N$ and $M_t$ the counting random variable of the marked Poisson Process at a given time $t$, the algorithm can be written as follows:
$$
\text{While } M_{T_\mn(k)} \leq \lceil - n_c N \log p \rceil, k = k+1
$$

\begin{propo}[Expectation of $\Niter$ in the sequential algorithm]
\eq{
\E{\Niter} & = & -N \log p + \sqrt{-2 N \log p \log n_c}
}

\begin{proofprop}
Let us write $t = - \log p$ and $m = \lceil n_c N t \rceil$; we have:
$$
\proba{ \Niter \leq k } = \proba{ M_{T_\mn(k)} \leq m} = \proba{ T_\mn(k) \geq T_m }
$$
Furthermore, $T_m = \Gamma_m / n_c N \sim \mathcal{N}(t,t / n_c N)$. Denoting by $P_\text{max}$ the random variable of the maximum of $n_c$ iid Poisson variable with parameter $Nt$, we have:
$$
\proba{ \Niter \leq k } \approx \proba{ T_\mn(k) \geq t } = \proba{ P_\text{max} \leq k }
$$
Finally, by doing as in the proof of proposition \ref{tpar_proba} the approximation $\mathcal{P}(Nt) \sim \mathcal{N}(Nt, Nt)$, $P_\text{max}$ becomes the maximum of $n_c$ iid Normal variables and we conclude using the extreme value theorem.
\end{proofprop}
\end{propo}
\subparagraph{Computing time of Moving Particles algorithm}
Since the expectations of the number of iterations are approximately the same for both algorithms (Algorithm \ref{algo_estim_quantile_2fold} and Algorithm \ref{algo_estim_quantile_seq}), we can finally derive the computing time of the quantile estimator as a function of the coefficient of variation:
\eq{ \label{comput_time_quant}
t_\text{par} & \approx & \dfrac{T}{n_c} \left( \dfrac{p \log p}{\delta q f(q)} \right)^2 \left( 1 + \dfrac{1}{T \log 1/p} +  \delta\gamma(q) \sqrt{2 n_c \log n_c } \right)
}
with $\gamma(q) = \dfrac{q f(q)}{-p \log p}$.

Eventually we can compare this time to the one from a naive Monte-Carlo estimator:
\eq{
\log \left( \dfrac{\tpar}{\tmc} \right) & = &  \log \left( T p (\log p)^2 \right) + \log \left(  1 + \dfrac{1}{T \log 1/p} + \delta \gamma(q) \sqrt{2 n_c \log n_c} \right) \\
}

\begin{rem}[Order of magnitude of $\gamma(q)$]
While there is no general result on the order of magnitude of $\gamma(q)$, it can be shown that in a lot of cases it remains small. For instance if one considers von Mises distributions, \cad distributions such that the \textit{cdf} $F$ has the following representation:
\eq{
1 - F(q) = \bar{F}(q) = c \exp \left( - \displaystyle\int_z^q \dfrac{1}{a(t)}\mathrm{d}t \right)
}
with $c$ a given positive constant and $a(\cdot)$ the auxiliary function of $F$: $a = \bar{F}/f$ (see \cite{embrechts1997modelling} Definition 3.3.18), one obtains:
$$
\dfrac{1}{\gamma(q)} = \dfrac{-p \log p}{q f(q)}  =  -\dfrac{a(q)}{q} \log c + \dfrac{a(q)}{q} \int_z^q \dfrac{\mathrm{d}t}{a(t)}
$$
In this equality, the first term tends to 0 (see \cite{embrechts1997modelling} Proposition 3.3.24) and the second one can be bounded from below by $1 - z/q$ for any $z \in \R \mid \forall x \in \R, x \geq z \Rightarrow a'(z) \geq 0$, which means $\gamma(q) \in [0, 1]$.

Furthermore for exponential tails, \cad $f(x) \underset{+\infty}{\sim} \exp ( - ( \frac{\mid x \mid}{x_0} )^k )$ for some $k \in \N$ and $x_0 \in \R^*_+$ one can write (see \cite{embrechts1997modelling} Example 3.3.23):
$$
p = -\dfrac{f(q)^2}{f'(q)} (1 + o(1)) = \dfrac{e^{-q^k}}{k q^{k-1}} (1 + o(1)) 
$$
which brings:
\begin{align*}
1/\gamma(q) &= \dfrac{-p \log p}{q f(q)} \propto \dfrac{1}{k} \left( 1 - \dfrac{\log(k q^{k-1})}{q^{k}} + o(\dfrac{1}{q^{k}}) \right) \underset{p \rightarrow 0}{\sim} \dfrac{1}{k}
\end{align*}
Eventually one can conclude that $\gamma(q)$ remains low.
\end{rem}
As for the probability estimator there is an extra term driven by $\sqrt{n_c \log n_c}$ and eventually the order of magnitude remains the same and the quotient $\tpar / \tmc$ is always driven by $p$ for \emph{small} values of $p$.

\section{Examples} \label{examples}

\subsection{Presentation of the examples}
\subsubsection{Watermarking detection \label{exemple_guy}}
This example is the one used by Cérou \textit{et al.} \cite{cerou2012sequential} and Guyader \textit{et al.} \cite{guyader2011simulation} to show the properties of their algorithms. Let $d \in \mathbb{N}^*$ be the dimension of the input space and $\U$ be a unit vector in $\mathbb{R}^d$; the failure domain is regarded as the interior of a double cone of axis $\U$ (see \cite{merhav2008optimal}):
\begin{equation}
F = \{ \mathbf{x} \in \mathbb{R}^d \mid \Phi(\mathbf{x}) = \dfrac{\mid \mathbf{x}^T \U \mid}{\parallel \mathbf{x} \parallel} > q \}
\end{equation}
The analytic relation between $p$ and $q$ writes as follows:
$$
p = \mathbb{P}(\Phi(\mathbf{X}) > q) = 1 - F(q) = 1 - G\left( \dfrac{(d-1) q^2}{1 - q^2} \right)
$$
with $F$ the \textit{cdf} of $\Phi(\mathbf{X})$ and $G$ the \textit{cdf} of a Fisher variable with $(1, d-1)$ degrees of freedom (see \cite{guyader2011simulation}).

\subsubsection{A two-degrees-of-freedom damped oscillator}
\begin{figure}
\centering
\includegraphics[scale=0.5]{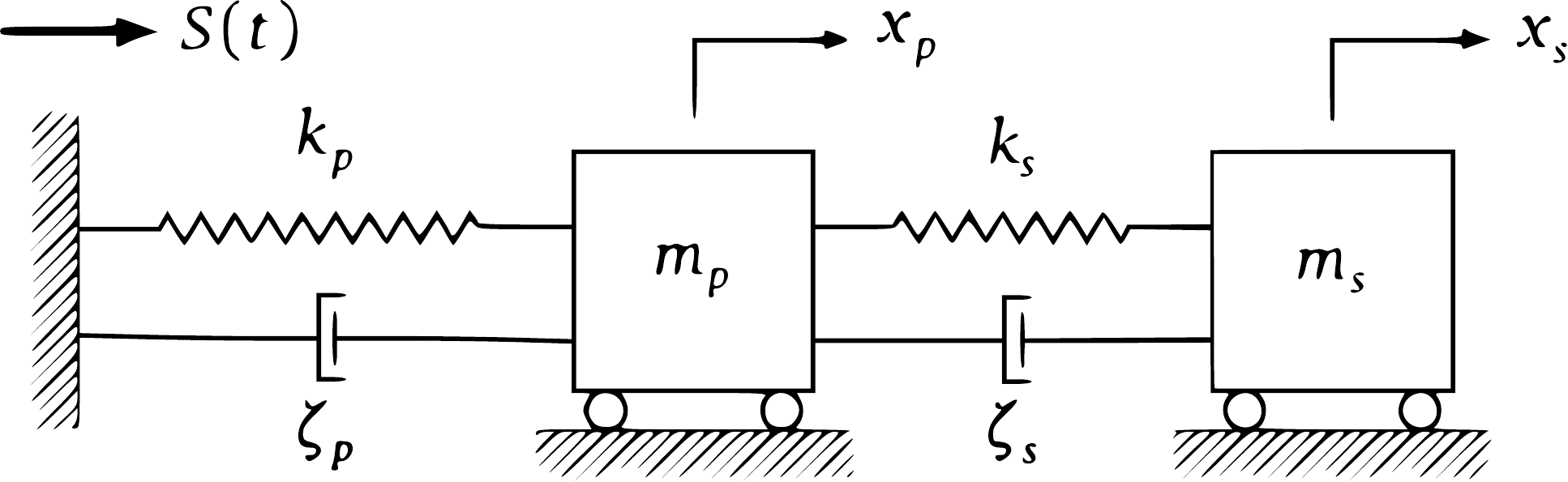}
\caption{{A 2 degrees of freedom damped oscillator (from \cite{dubourg2011metamodel})}}
\label{2dof_schema}
\end{figure}

This example sketched in Figure \ref{2dof_schema} was first proposed by Der Kiureghian and De Stefano \cite{kiureghian1991efficient} and then used by Bourinet \textit{et al.} \cite{bourinet2011assessing} and Dubourg \cite{dubourg2011metamodel}.

It is a two degrees of freedom damped oscillator characterised by masses $m_p$ and $m_s$, spring stiffnesses $k_p$ and $k_s$, natural frequencies $\omega_p^2 = k_p / m_p$ and $\omega_s^2 = k_s / m_s$ and damping ratios $\zeta_p$ and $\zeta_s$. Igusa and Der Kiureghian \cite{igusa1985dynamic} showed that the mean-squared relative displacement of the secondary spring under a white noise base acceleration with intensity $S_0$ writes as follows:
$$
\mathbb{E}[x_s^2] = \pi \dfrac{S_0}{4 \zeta_s \omega_s^2} \dfrac{\zeta_a \zeta_s}{\zeta_p \zeta_s (4 \zeta_a^2 + \theta^2) + \gamma \zeta_a^2} \dfrac{(\zeta_p \omega_p^3 + \zeta_s \omega_s^3) \omega_p}{4 \zeta_a \omega_a^4}
$$
with $\gamma = m_s/m_p$, $\omega_a = (\omega_p + \omega_s)/2$, $\zeta_a = (\zeta_p + \zeta_s)/2$ and $\theta = (\omega_p - \omega_s)/\omega_a$

Finally, Der Kiureghian and De Stefano \cite{kiureghian1991efficient} showed that the \lsf could write under reasonable approximation as follows:
\begin{equation}
g(\mathbf{x}) = F_s - p \, k_s \, \sqrt{\mathbb{E}[x_s^2]}
\end{equation}
with $F_s$ the force capacity of the secondary spring and $p$ a peak factor here set to 3 as in \cite{dubourg2011metamodel}.

Table \ref{2dof_table} presents the probabilistic model used.
\begin{table}[!h]
\centering
\renewcommand{\arraystretch}{1}
\begin{tabular}{|c|c|c|}
\hline
Variable & Mean & CV (\%) \\
\hline
$m_p$ & $1.5$ & $10$ \\
$m_s$ & $0.01$ & $10$ \\
$k_p$ & $1$ & $20$ \\
$k_s$ & $0.01$ & $20$ \\
$\zeta_p$ & $0.05$ & $40$ \\
$\zeta_s$ & $0.02$ & $50$ \\
$F_s$ & $\{15 ; 21.5 ; 27.5 \}$ & $10$ \\
$S_0$ & 100 & 10 \\
\hline
\end{tabular}
\caption{Stochastic model of the oscillator}
\label{2dof_table}
\end{table}
As it uses lognormal distributions and reversible kernel is defined in the standard space a conversion is done before each call to the limit-state function.

\subsection{Estimation of failure probability}

\paragraph{Watermarking detection}
Here we set $d=20$, $q = 0.95$ and we try to estimate $p=1 - G\left( \dfrac{(d-1) q^2}{1 - q^2} \right) = 4.704 \, 10^{-11}$.

\paragraph{A two-degrees-of-freedom damped oscillator}
Failure is defined as $g(\x) < 0$. There is no analytical expression available to benchmark our algorithm and so reference values have been calculated with reference estimators as described Table \ref{2dof_ref}.
\begin{table}
\centering
\renewcommand{\arraystretch}{1}
\begin{tabular}{|c|c|c|c|c|}
\hline
$F_s$ & Method & Nbr of points & Proba & CV \\
\hline \hline
$15$ & Monte-Carlo brut & $2000000$ & $4.8015 \, 10^{-3}$ & $0.01018$ \\
$21.5$ & Monte-Carlo brut & $10000000$ & $4.34 \, 10^{-5}$ & $0.048$ \\
$27.5$ & Subset Simulation & $4000000$ & $3.745 \, 10^{-7}$ & $0.0286$ \\
\hline
\end{tabular}
\caption{Reference values for the two-degrees-of-freedom damped oscillator}
\label{2dof_ref}
\end{table}

\subsubsection{Effect of the choice of $N$ and $n_c$ for a total of $1000$ particles }\label{effet_N_nc_proba}
\begin{figure}[!ht]
%\begin{minipage}[T]{0.5\textwidth}
\centering
\subfloat[Watermarking detection]{\includegraphics[width=0.45\textwidth]{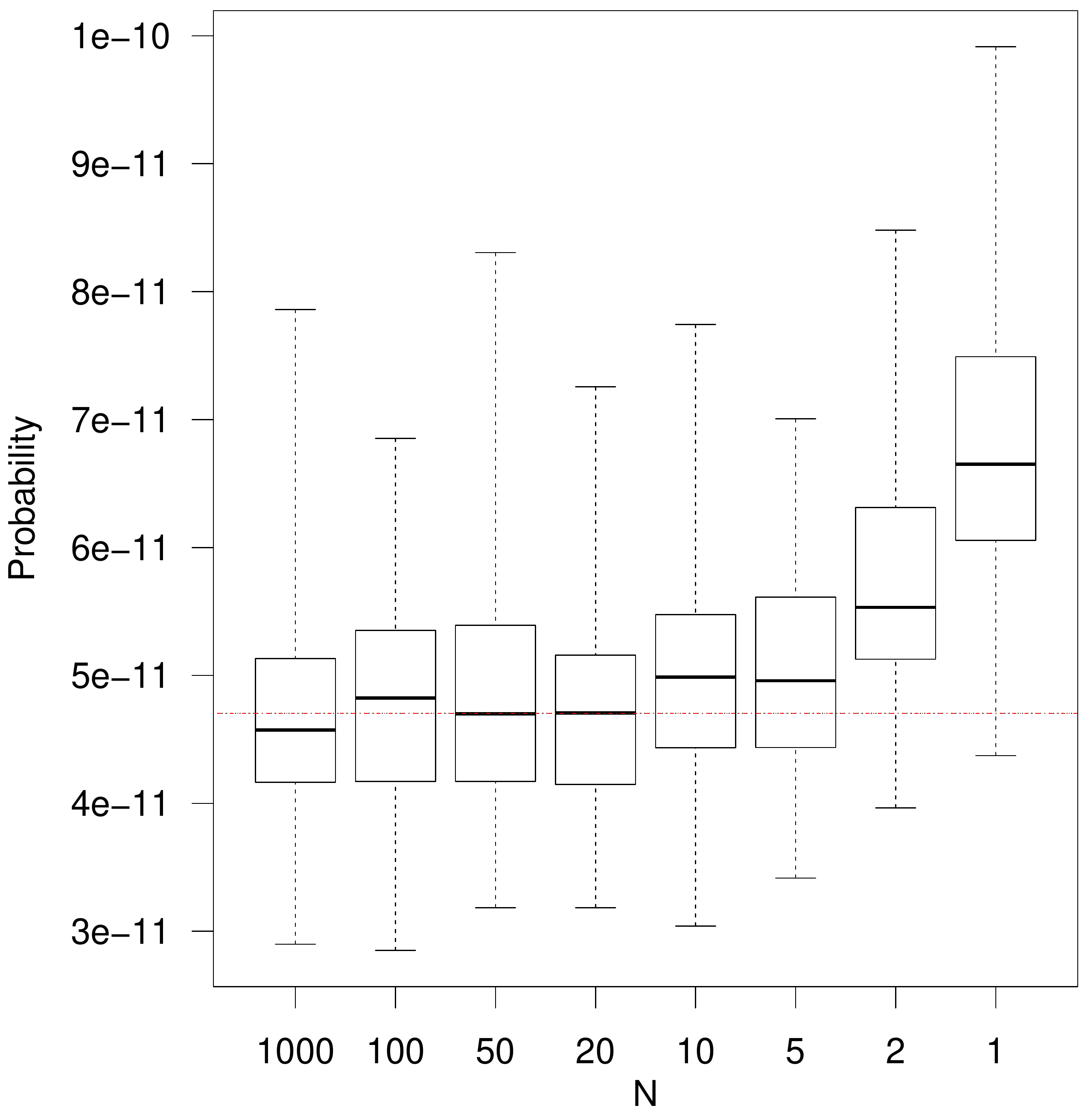}}
\subfloat[2 d-o-f oscillator with $\E{F_s} = 15$]{\includegraphics[width=0.45\textwidth]{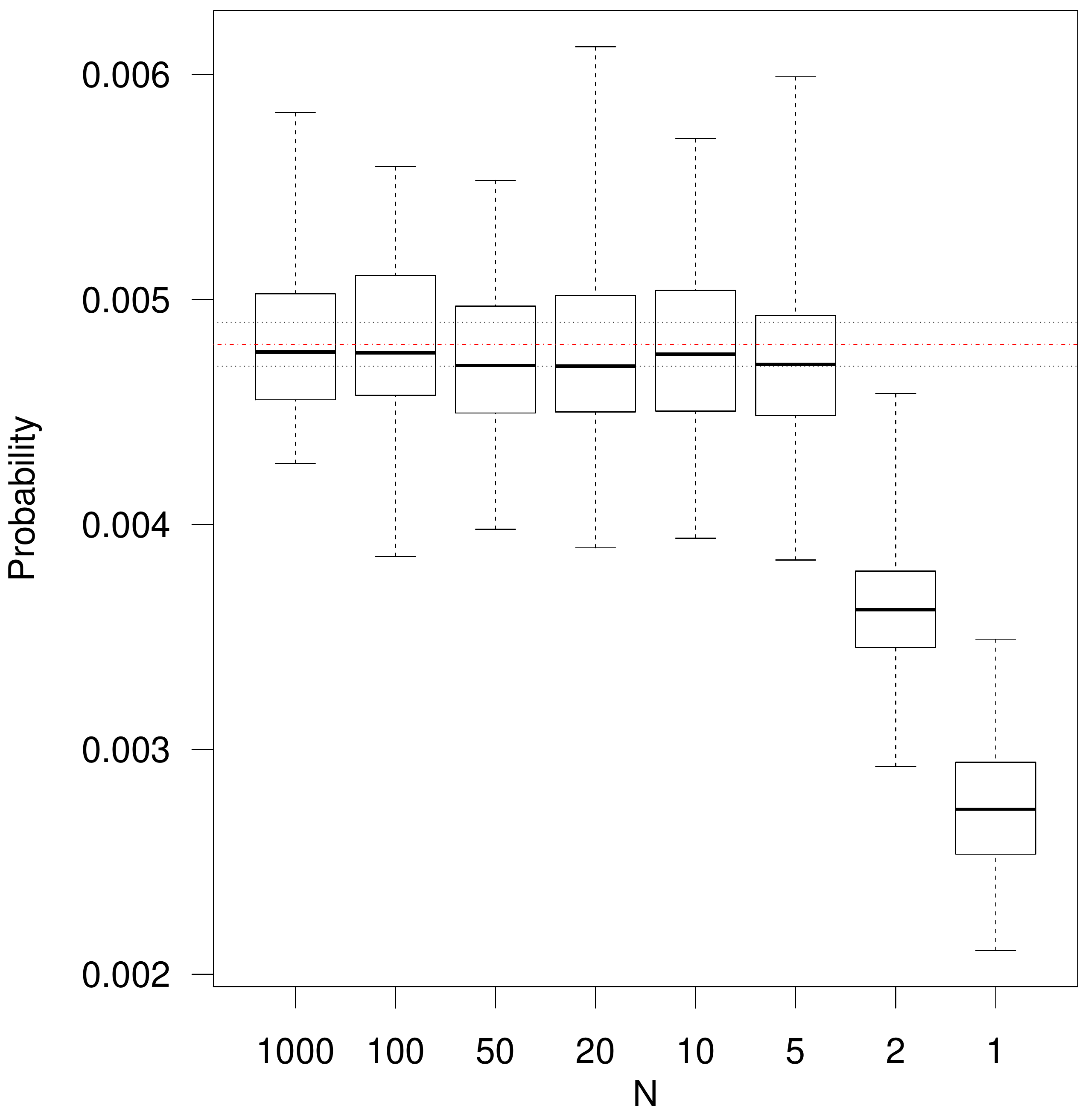}}\\
%\end{minipage}
%\begin{minipage}[T]{0.5\textwidth}
%\centering
\subfloat[2 d-o-f oscillator with $\E{F_s} = 21.5$]{\includegraphics[width=0.45\textwidth]{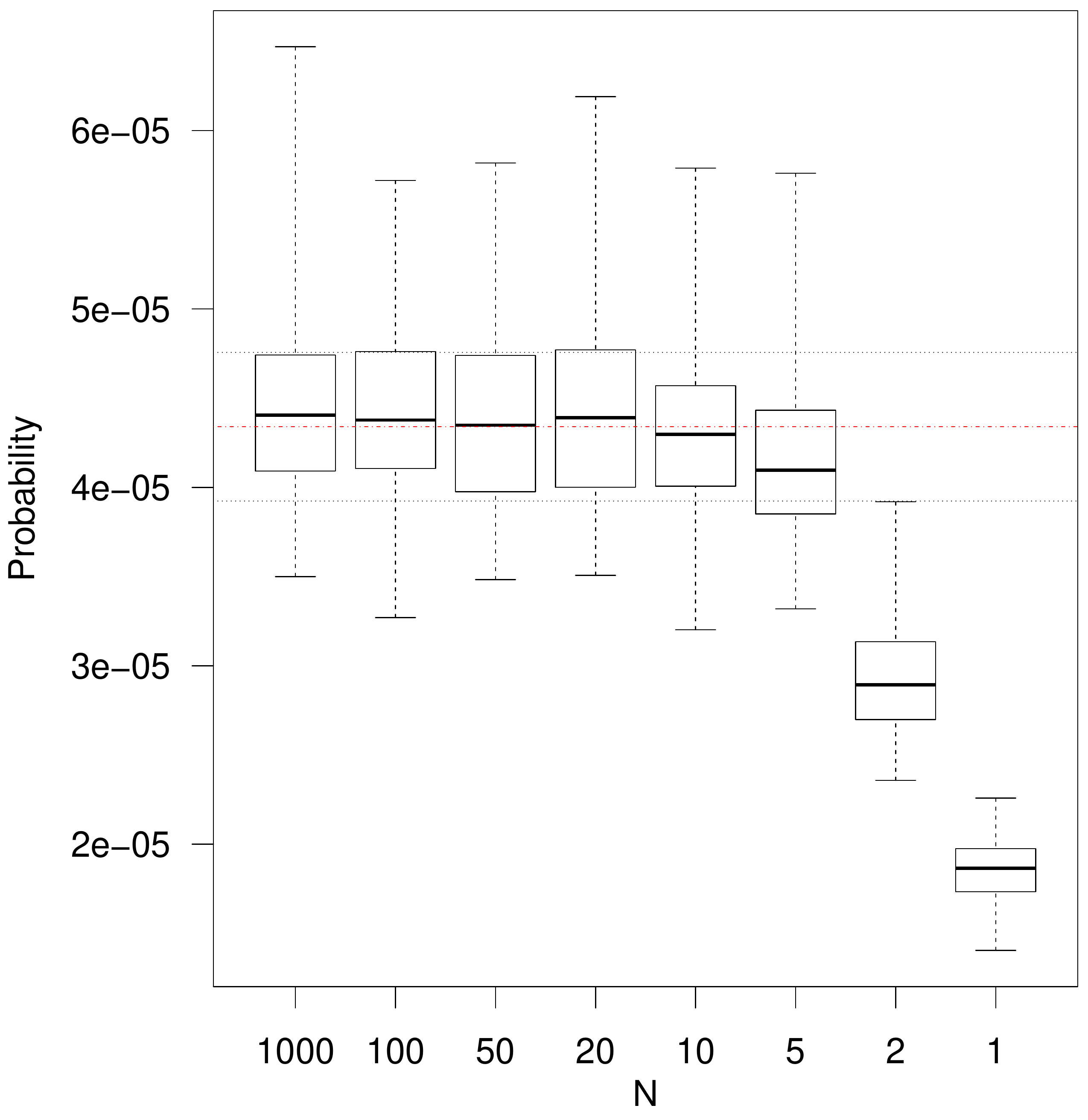}}
\subfloat[2 d-o-f oscillator with $\E{F_s} = 27.5$]{\includegraphics[width=0.45\textwidth]{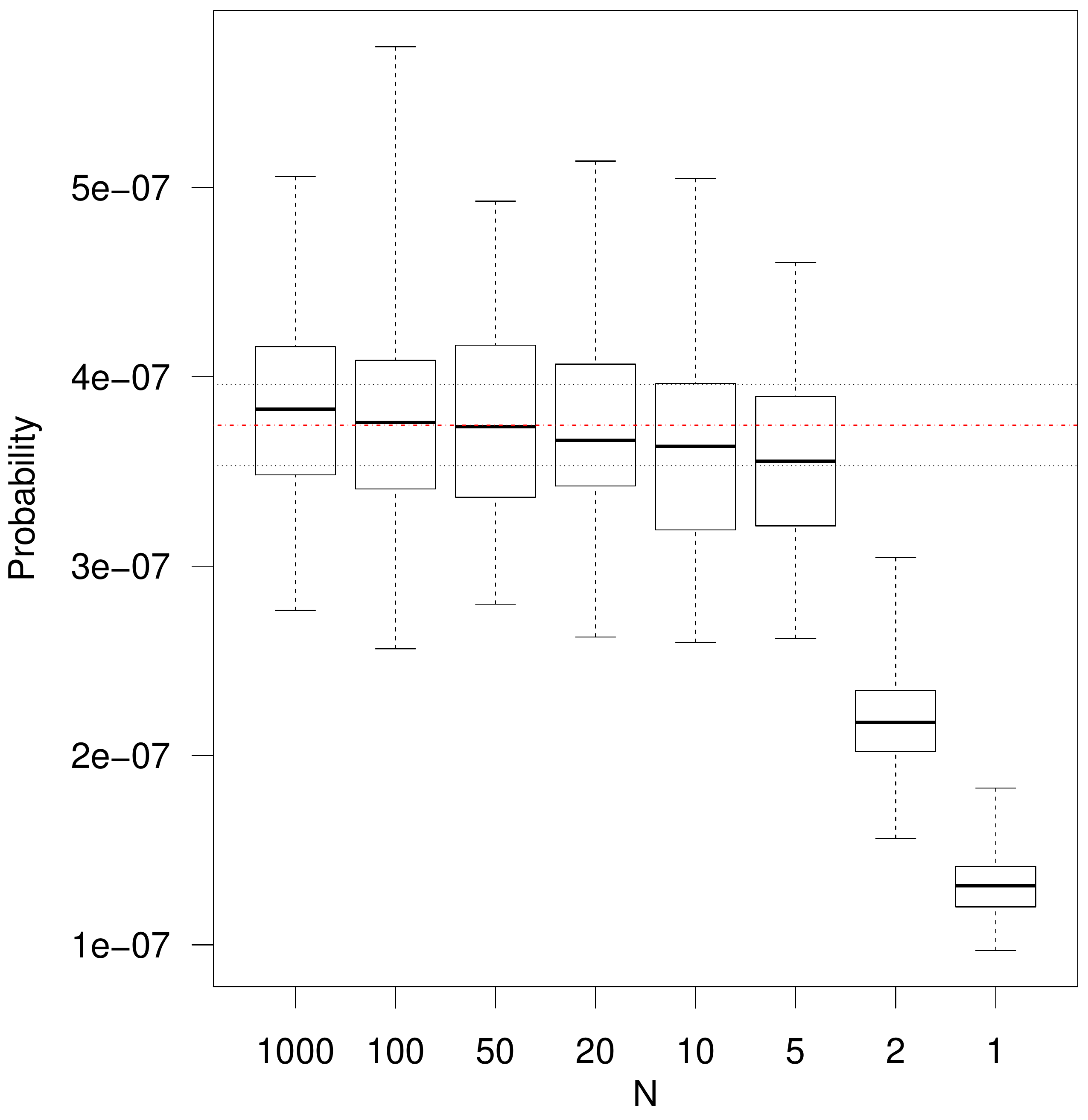}}
%\end{minipage}
\caption{Boxplots of the probability estimator over 100 simulations}
\label{boxplots_proba_est}
\end{figure}

The purpose of this part is to precise the minimal acceptable value for $N$ as the smaller $N$ the faster the whole algorithm. The following configurations have been tested, expressed as "$n_c \times N$": "1x1000", "10x100", "20x50", "50x20", "100x10", "200x5", "500x2" and "1000x1". Results are shown in Figure \ref{boxplots_proba_est} as boxplots of 100 simulations, whiskers extending to the extreme values. The reference value is displayed with the red dashed line, and in the case of the 2 d-o-f oscillator 95\% confidence interval is displayed as well with the black dashed lines (see Table \ref{2dof_ref}).
For all examples it appears that for $N = 10$ and above the estimations are almost the same.

\subsubsection{Effective computing time}

In this section we look at the effective computing time of the estimators, \cad the maximum number of calls to the \lsf made by each of the $n_c$ algorithms for a given configuration. Especially we intend to check the consistency of formula \eqref{comput_time_proba} while number of calls not only depends on the \textit{burn-in} parameter $T$ and the number of mutations, but also the number of rejected mutations, \cad transitions where the starting point was a replica of the moving particle and all transitions were refused (see \ref{subsection_practical_implementation_algo_par}); this situation especially arises when the number of particles gets small ($N \leq 5$).

\begin{figure}[!ht]
%\begin{minipage}[T]{0.5\textwidth}
\centering
\subfloat[Watermarking detection]{\includegraphics[width=0.45\textwidth]{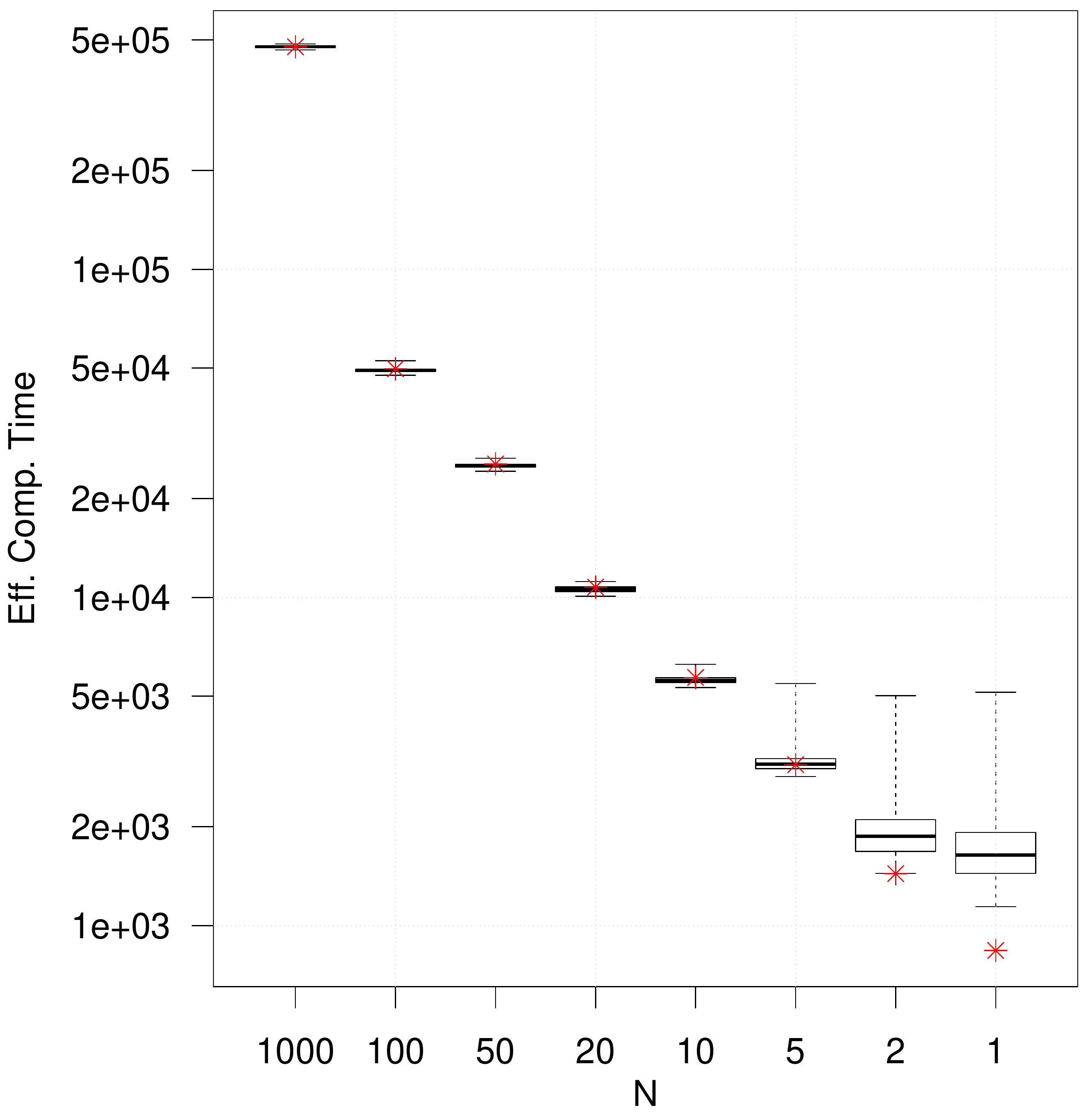}}
\subfloat[2 d-o-f oscillator with $\E{F_s} = 15$]{\includegraphics[width=0.45\textwidth]{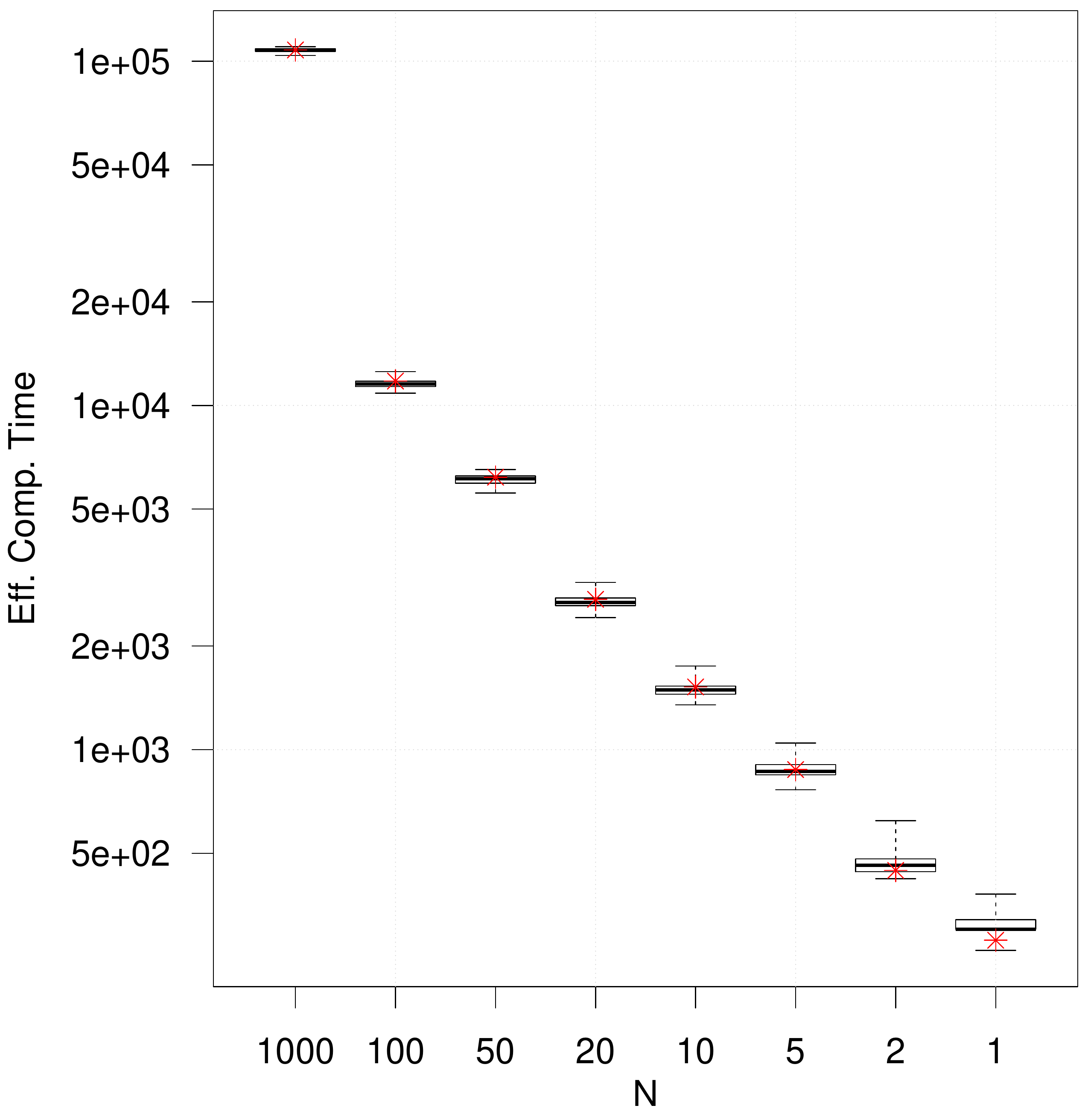}}\\
%\end{minipage}
%\begin{minipage}[T]{0.5\textwidth}
%\centering
\subfloat[2 d-o-f oscillator with $\E{F_s} = 21.5$]{\includegraphics[width=0.45\textwidth]{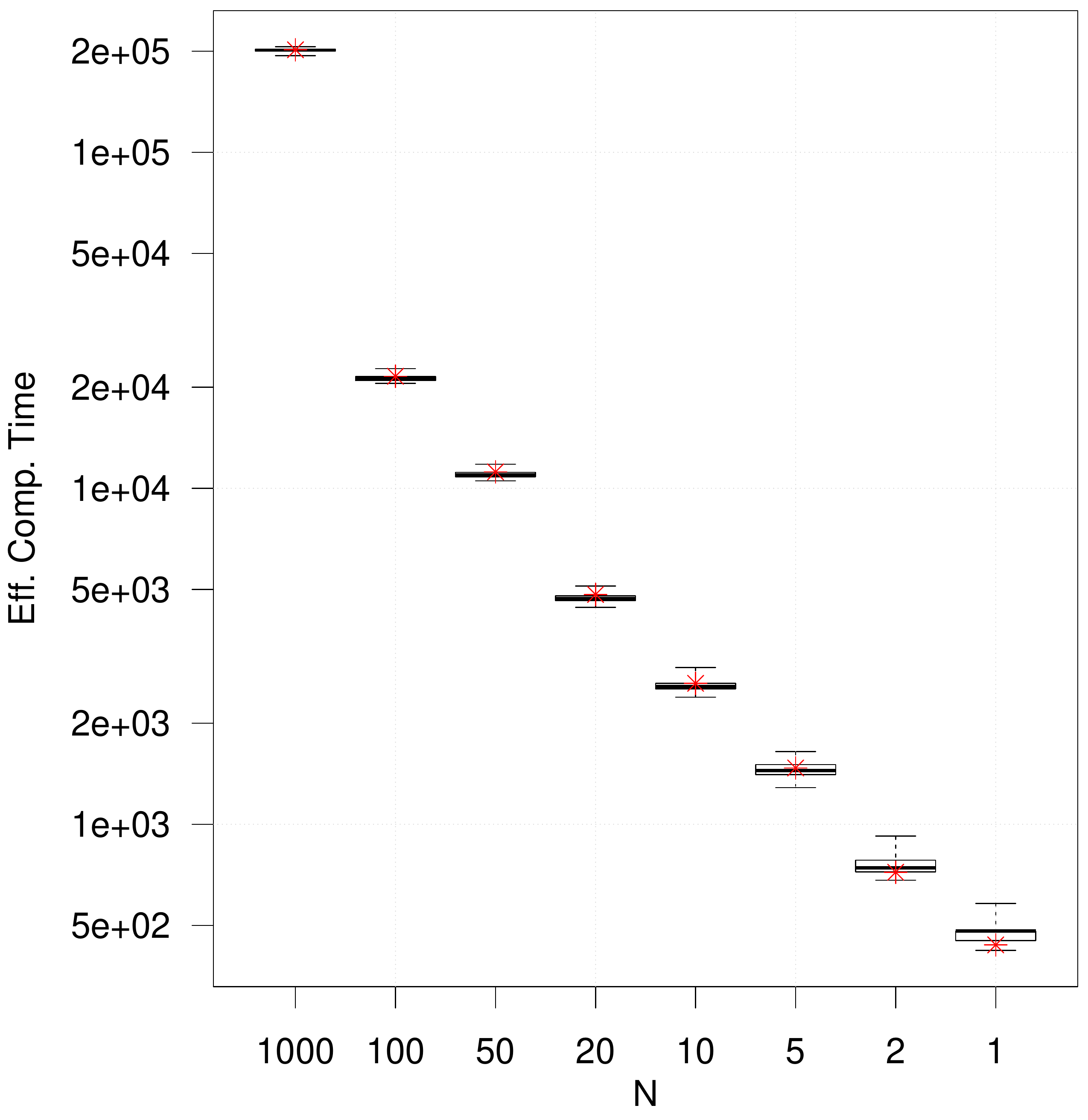}}
\subfloat[2 d-o-f oscillator with $\E{F_s} = 27.5$]{\includegraphics[width=0.45\textwidth]{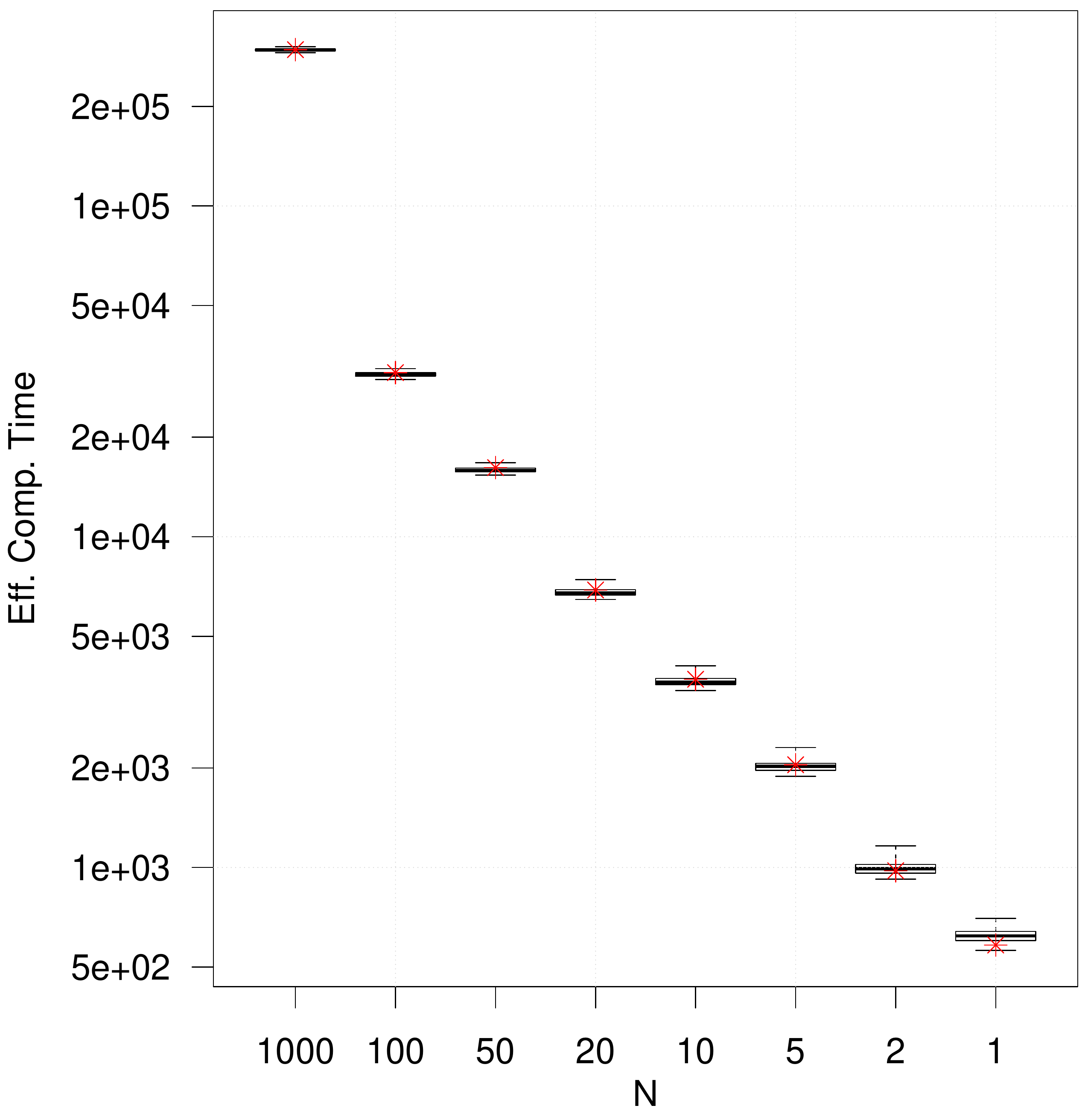}}
%\end{minipage}
\caption{Effective computing time of the probability estimator over 100 simulations}
\label{boxplots_proba_time}
\end{figure}

As for the probability, results are displayed as boxplots overs 100 simulations in Figure \ref{boxplots_proba_time}. The red dots show the theoretical value given by equation \eqref{comput_time_proba}.
One can see that apart from the two last configurations of watermarking detection examples, these values are in good agreement with the empirical results.

\subsubsection{Conclusion on the estimator}
The moving particle point of view allows us to define a new estimator for extreme probabilities. This estimator can be related to Multilevel Splitting methods and especially had already been proposed in this framework by Guyader \textit{et al.} \cite{guyader2011simulation}. Nevertheless we can now compute it in a parallel way which makes it more efficient than any Multilevel Splitting strategy; especially gain is approximately of $50\%$ in terms of computing time for an equivalent precision comparing to \textit{Subset Simulation} as described by \cite{au2001estimation}.
Comparing to naive Monte-Carlo, a standard choice of $T = 20$ for the \textit{burn-in} parameter makes our algorithm better as soon as $p \lesssim 10^{-3}$ and gain increases as $1/p$.

%All these results are summed up table \ref{tableau_comp_estim_proba}.
%\begin{table} [!ht]
%\centering
%\renewcommand{\arraystretch}{2.2}
%\begin{tabular}{|c|c|c|c|}
%\hline \hline
%Method & $\delta^2$ & Eff. comp. time \\
%\hline \hline
%Naive Monte-Carlo & $\dfrac{1}{Np}$ & $\dfrac{1}{\delta^2 n_c p}$\\
%Subset Simulation \cite{au2001estimation} & $T  \dfrac{(1+\gamma)}{N p_0} \left( \dfrac{\log p}{\log p_0} \right)^2$ & $\lceil \dfrac{1}{n_c}  \dfrac{T(1+\gamma)}{(\log p_0)^2 p_0} \left( \dfrac{\log p}{\delta} \right)^2 \rceil$\\
%Sequential Monte-Carlo \cite{cerou2012sequential} & $T  \dfrac{1}{N p_0} \left( \dfrac{\log p}{\log p_0} \right)^2$ & $\lceil \dfrac{1}{n_c}  \dfrac{T}{(\log p_0)^2 p_0} \left( \dfrac{\log p}{\delta} \right)^2 \rceil$\\
%Guyader \textit{et al.} \cite{guyader2011simulation} & $T  \dfrac{(\log p)^2}{N}$ & $\lceil T \left( \dfrac{\log p}{\delta} \right)^2 \rceil$ \\
%Moving particules & $T  \dfrac{(\log p)^2}{N}$ & $\lceil \dfrac{T}{n_c} \left( \dfrac{\log p}{\delta} \right)^2 \rceil$\\
%\hline
%\end{tabular}
%\caption{Comparison of moving particles algorithm with usual strategies for estimating a probability \label{tableau_comp_estim_proba}}
%\end{table}
Practically speaking, the minimal number of particles to be considered in each algorithm \ref{algo_atteinte_defaillance_par} seems to depend on the limit-state function and taking $N \geq 10$ appears to be a conservative choice.

\subsection{Estimation of quantiles}

We use the example presented section \ref{exemple_guy} to qualify our algorithm. Unlike probability estimator, which was indeed the same as Guyader \textit{et al.} one's (we only intended to check parallelisation capacity) we have presented here above a new estimator. We thus set $p = 4.704 \, 10^{-11}$ and try to find back $q = 0.95$.

\subsubsection{Effect of the choice of $N$ and $n_c$ for a total of $1000$ particles}
\begin{figure}[!ht]
%\begin{minipage}[T]{0.5\textwidth}
\centering
\subfloat[Quantile estimates]{\includegraphics[width=0.45\textwidth]{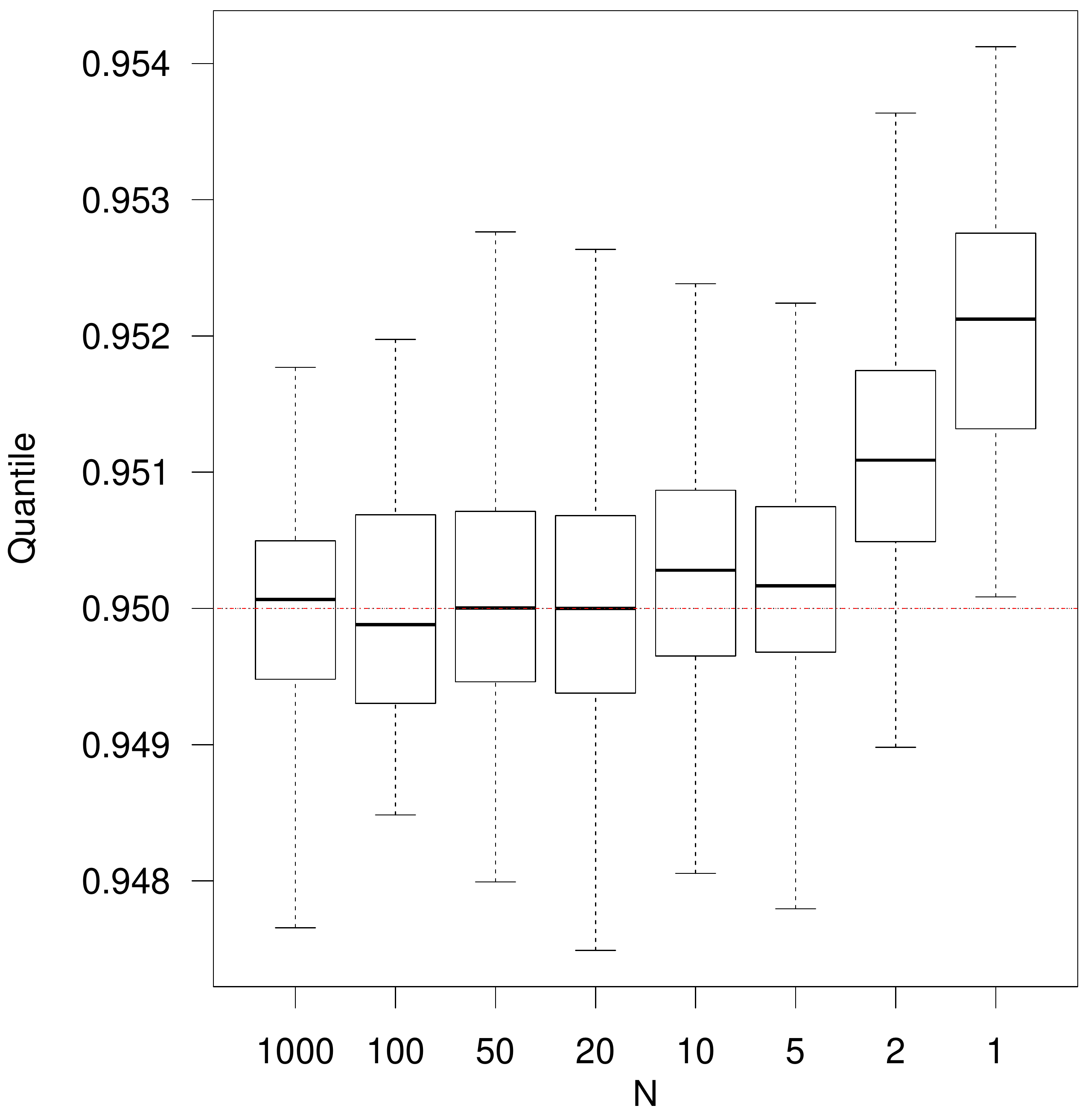} \label{q_phi}}
\subfloat[Effective computing time]{\includegraphics[width=0.45\textwidth]{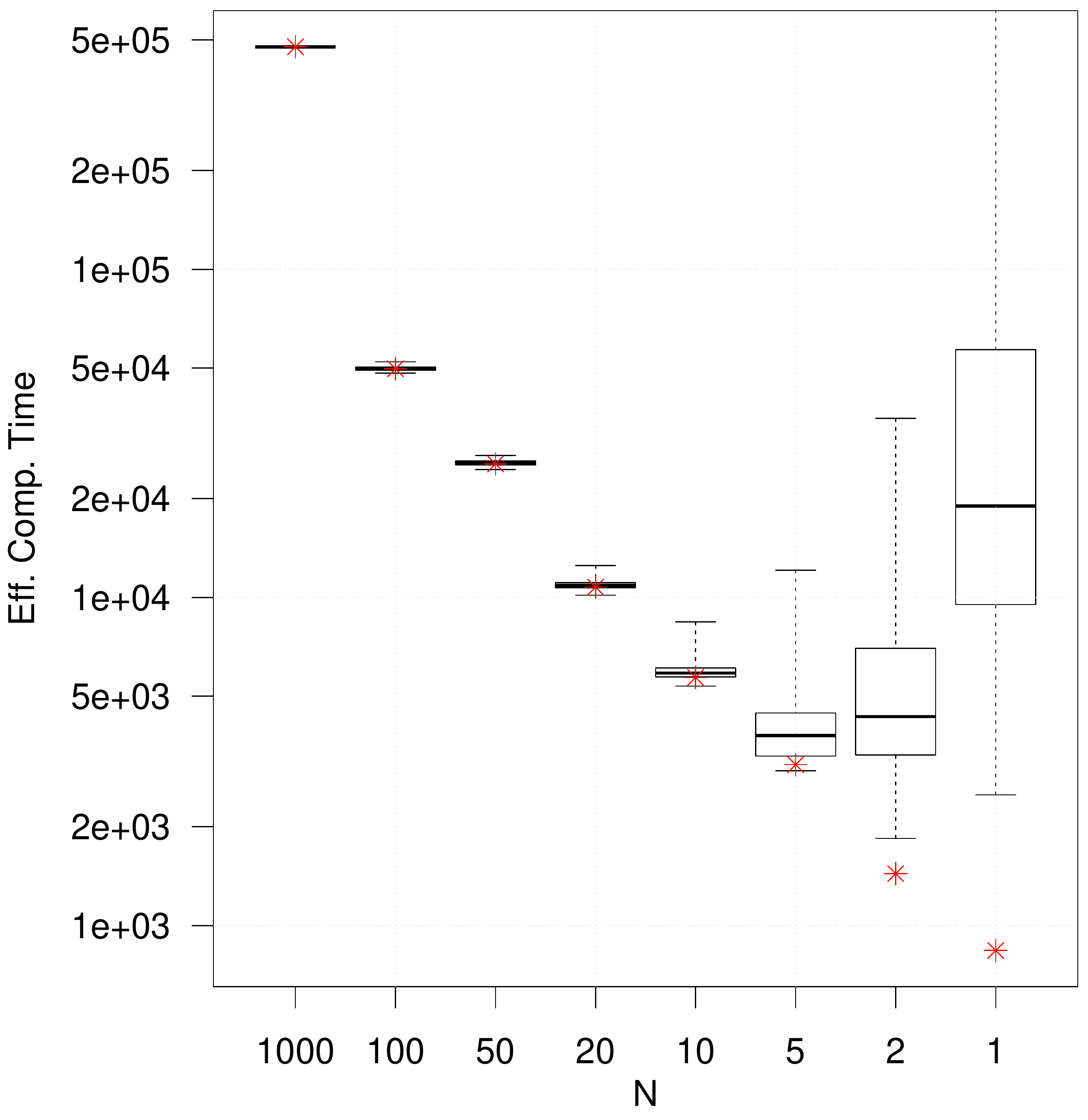} \label{time_q_phi}}\\
%\end{minipage}
%\begin{minipage}[T]{0.5\textwidth}
%\centering
\subfloat[Total number of events]{\includegraphics[width=0.45\textwidth]{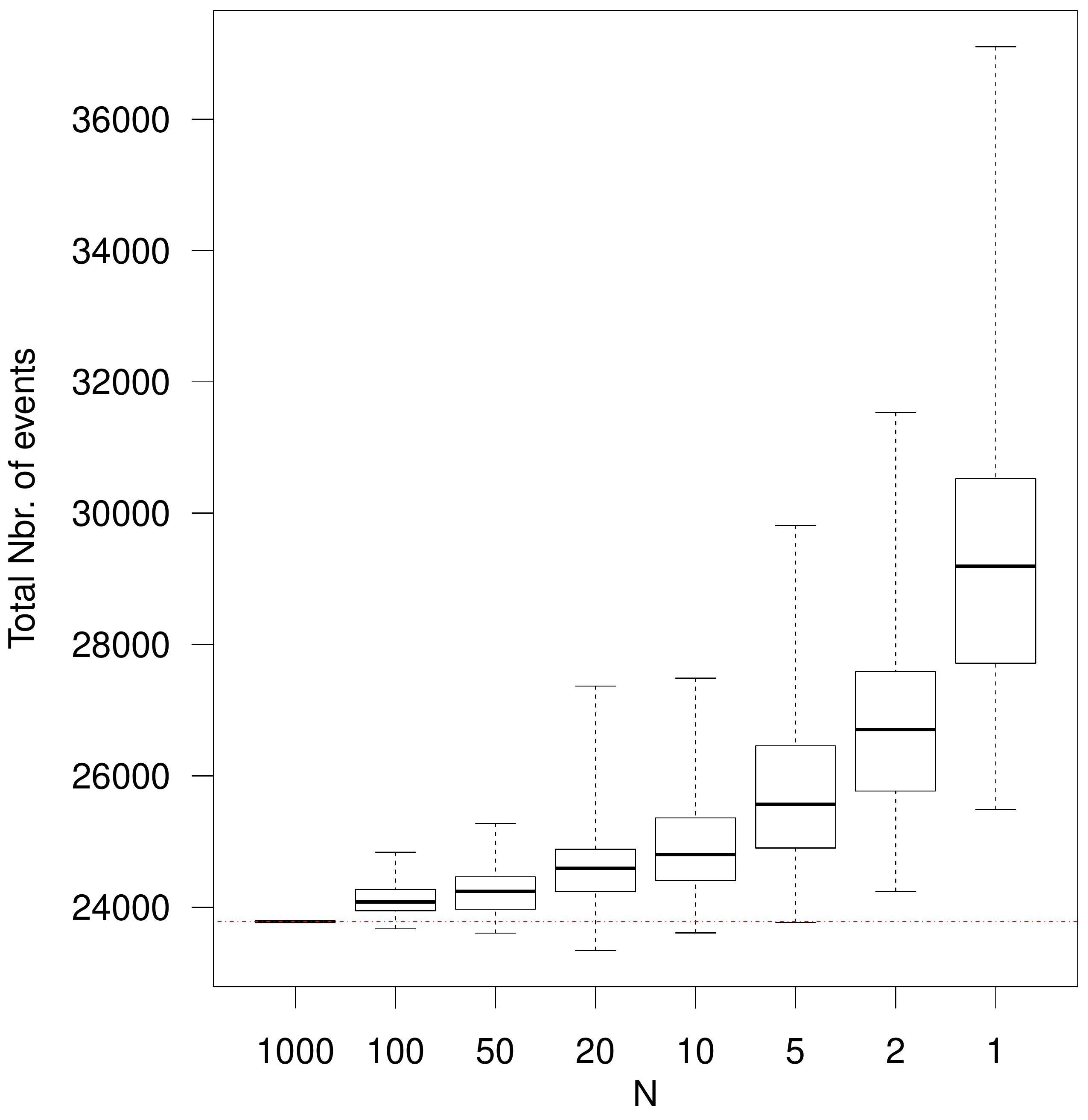} \label{event_q_phi}}
\subfloat[Empirical failure]{\includegraphics[width=0.45\textwidth]{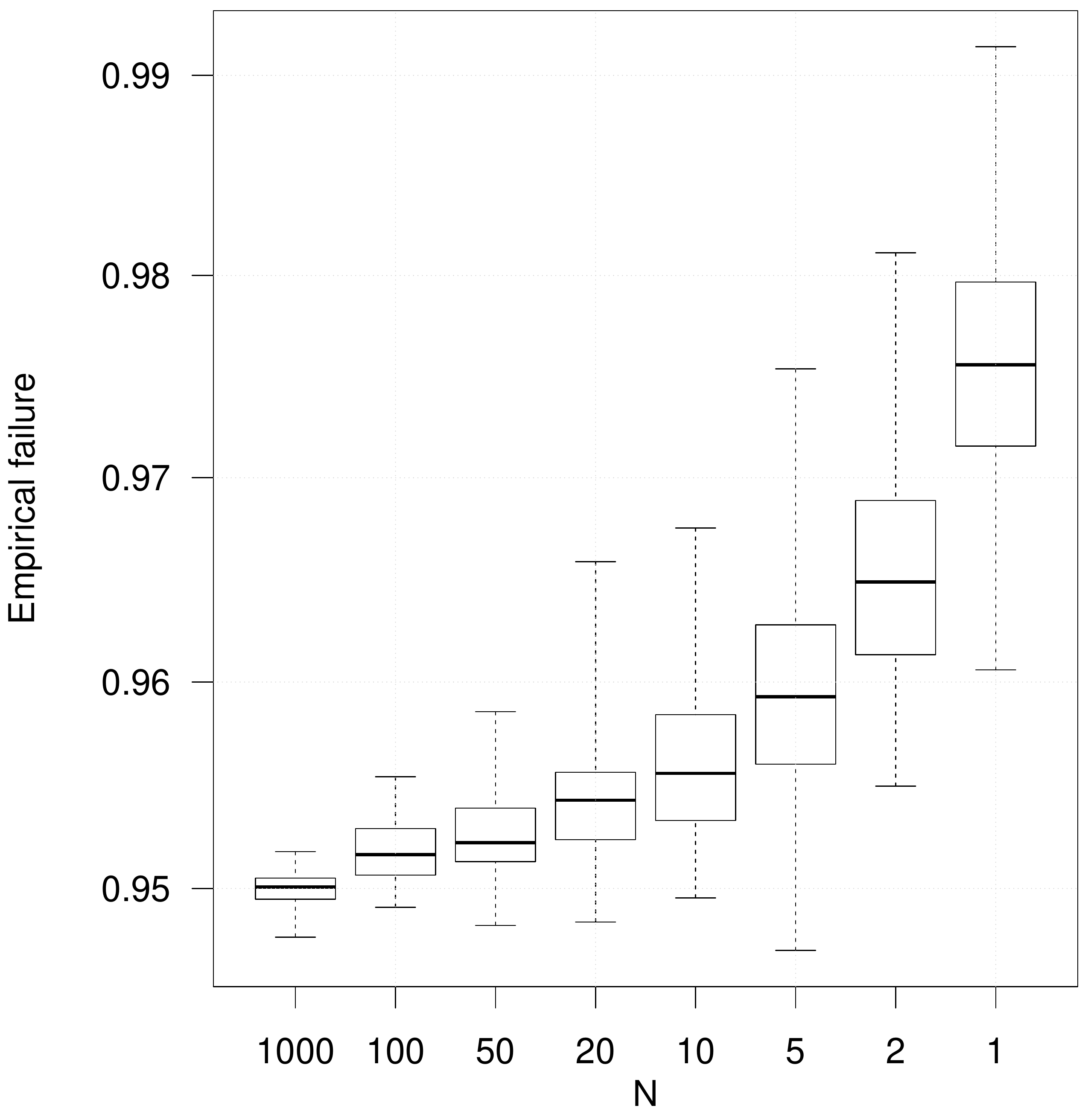} \label{failure_q_phi}}
%\end{minipage}
\caption{Statistics on quantile estimator over 100 simulations}
\label{q_est_figure}
\end{figure}

The context is the one presented section \ref{effet_N_nc_proba}. Results are displayed in Figure \ref{q_est_figure} as boxplots over 100 simulations, whiskers extending to the extreme values and reference value is added to the plot with a dashed line.
Once again for $N \geq 10$ the estimations seem to be almost the same. One could expect to get the same type of results as for the probability because the algorithm are intrinsically the same, \cad a move of particles. Thus a higher estimation of the probability means a too low number of iterations, \cad that particles are moving \emph{too fast} which will directly produce an overestimation of the quantile.

\subsubsection{Effective computing time}
As for the probability estimator we intend to validate formula \eqref{comput_time_quant}. The results are presented in Figure \ref{time_q_phi} and show a good agreement with the formula apart from the \emph{extreme} cases ($N \leq 5$). This is because the smaller the population, the higher the intermediate failure level (cf. Figure \ref{failure_q_phi}) and so greater the number of transitions to stop the algorithm. As for small populations ($N \leq 5$) the generation step does not work well, more transitions means more generation steps with the rejection of all transitions, \cad "useless" calls to the limit-state function.

We also check the total number of mutations as this should be ideally equal to the targeted one's: $m = \lceil -n_c N \log p \rceil$ and in practice as close as possible. Especially we have accepted here to take a risk $\alpha = 5\%$ (see equation \eqref{choix_m0}) not to have enough sample at the end of the algorithm.
In Figure \ref{event_q_phi} we can see that in some cases the algorithm did not produce enough events. Especially the number of "too short" algorithms are 5, 4, 3, 3 and 1 respectively. On a total of 100 simulations this is in good agreement with the parameter $\alpha$ set to $5\%$.
\subsubsection{Conclusion on the estimator}
Unlike the probability estimator, which was indeed the one proposed by Guyader \textit{et al.}, the moving particles point of view allows us to define a new estimator with a reduced bias. Statistical properties are only asymptotic as for naive Monte-Carlo estimator but confidence intervals can be computed without any estimation of the \textit{pdf}. In terms of computing time, it has the same asymptotic properties as the estimator of probabilities. As far as we know, it is the first parallel alternative to naive Monte-Carlo and thus allows for a massive gain for quantile estimation.

\section{Application to the construction of first Design of Experiments} \label{section_DoE}

In this section we intend to apply the moving particles point of view to the building of DoE for meta-model based algorithms. Especially it means we only keep the \emph{moving part} while dropping the constraint on \emph{exact} sampling.

\subsection{Meta-modelisation and first DoE}
\paragraph{The fixed-time framework}

While Algorithm \ref{algo_estim_proba} seen above is optimal compared to Multilevel Splitting methods, it still requires a lot of calls to the limit-state function. Furthermore this number of calls directly depends on the unknown parameter $p$ while in a practical context it is more often about using a given number of calls in the best way.

In this scope meta-model based algorithms use this computational budget to fit a surrogate model and then use it instead of the true function to estimate probabilities or quantiles.
 
\paragraph{The importance of the first DoE}

A major issue with these techniques is the control of the \emph{fidelity} of the surrogate model to the true function; especially one have to make sure the input space is explored enough to get a \emph{good} approximation of the failure domain. The sampling strategies are usually twofold: first an initial sampling which aims at giving an overall knowledge of the function, then an iterative refinement of the surrogate model according to a given criterion. The \emph{quality} of the first DoE is of great importance as it is highly unlikely that the refinement step be able to find back the boundary between safety and failure domain if no failing points are known. Furthermore the computing time constraint does not allow for a revision of the strategy if unsuccessful.

\paragraph{What is it to learn about}

It is noteworthy that regardless of the kind of meta-model techniques used (classification through Support-Vector Machine like Bourinet \textit{et al.} algorithm $^2\text{SMART}$  \cite{bourinet2011assessing} or regression as in AKMCS \cite{echard2011ak} or MetaIS \cite{dubourg2011metamodel}), the meta-model is always \emph{finally} used as a classification, directly or through the probability of being in the failure domain (MetaIS). Thus there is no real need to get a good knowledge of the limit-state function over the whole input space but only "around" the boundary delimiting $F$: it does not matter if regression is accurate in safety and failure domains, the prediction only needs to belong to the right domain.

Nevertheless usual strategies of \textit{Space filling} do not take benefit from this analysis and look at a global learning of the limit-state-function in the input space. Unfortunately this brings to direct exponential dependency on the dimension of the input space. On the other hand density-based DoEs (like in AKMCS) use the distribution of the input variable to sample from and select some points by clustering. While it does not depend on the dimension of the input space anymore, it is very unlikely to produce failing point as by definition $p$ is supposed to be small. An ideal strategy would indeed produce only pairs of points across the boundary; there is no need to get a sharp estimation of $g$ on both domains.

These observations lead us to propose a new strategy for first DoE based on algorithm \ref{algo_atteinte_defaillance} as it allows to get failing points \emph{quickly}. 

\subsection{Adaptation of algorithm \ref{algo_atteinte_defaillance} to the construction of first DoE}

\paragraph{First DoE}
As pointed out Section \ref{subsection_practical_implementation_algo_par} we can afford an approximation of the conditional sampling if we only intend to \emph{move}. The first things we need is to define a meta-model with a minimal-sized DoE, then we will be able to generate as many Markov Chains as desired number of failing points. Usual techniques use Maximum Likelihood Estimation to estimate the hyper-parameters of a meta-model, which requires at least $d+1$ points, given $d$ the dimension of the input space (for dimensional ranges and standard deviation). We thus expect a final number of calls to the limit-state-function:
\begin{equation} \label{Ncall_MAPES}
N_\text{final} \approx (d + 1) + N_\text{fail}  (- \log p)
\end{equation}
with $N_\text{fail}$ the number of failing points desired and so a linear dependency on the dimension. Finally we choose here to use Gaussian Process Regression (Kriging) and without any expert knowledge we fix the trend to be equal to the failure threshold: estimating it would implicitly exclude some unknown parts of the input space as this estimation done with the data will be lower than $q$.

\begin{algo}[Algorithm for getting a first DoE] \label{algo_MAPES} \ 
\begin{enumerate}
\item Sample $d+1$ points according to $\mu^X$ and calculate $g$
\item Learn a first meta-model $\widetilde{g}$ with $trend = q$
\item Do $N_\text{fail}$ times
	\begin{itemize}
	\item Sample $\mathbf{X}_1$ with respect to $\mu^X$
	\item Evaluate $g$: $g(\X_1) = y_1$; $m = 1$
	\item Train the meta-model
	\item While $y_m < q$
		\begin{itemize}
		\item $\X_{m+1} = \X_m$; $y_{m+1} = y_m$
		\item Do $T$ times
		\begin{itemize}
		\item[+] $\mathbf{X}^* \sim K(\X_{m+1}, \cdot)$
		\item[+] $\widetilde{g}(\X^*) = y^*$
		\item[+] If $y^* > y_{m+1}$, $y_{m+1} = y^*$ and $\X_{m+1} = \X^*$
		\end{itemize}
		\item Evaluate $g$: $g(\X_{m+1}) = y_{m+1}$
		\item Train the meta-model
		\item If $y_{m+1} < y_m$, $\X_{m+1} = \X_m$ and $y_{m+1} = y_m$
		\item $m = m+1$
	\end{itemize}
	\end{itemize}
\end{enumerate}
\end{algo}
One could notice that the \textit{burn-in} parameter $T$ is not exactly a proper \textit{burn-in} any more because the current threshold changes during the loop. As the main goal here is to move to the failure domain, there is a trade-off between huge moves and keeping close to the already explored input space. In fact this \emph{pseudo burn-in} allows for a move to "the end of the information".

\subsection{Exemples \label{MAPES_exemples}}

We test our algorithm on the previous limit-state functions and on some others presented below. We focus on the number of calls to the limit-state-function which should be distributed according to equation \eqref{Ncall_MAPES}. For all examples, $10$ failing points were asked to the algorithm and $T$ is set to 20.

\paragraph{A two-dimensional four branches serial system}
This example has been originally proposed by P.-H. Waarts \cite{waarts2000structural} and is defined as follows in the standard space:
\begin{equation}
g : \mathbf{x} \in \mathbb{R}^2 \longmapsto \min \left(
3 + \dfrac{1}{10} (x_1 - x_2)^2 - \dfrac{1}{\sqrt{2}} \mid x_1 + x_2 \mid, \dfrac{7}{\sqrt{2}}- \mid x_1 - x_2 \mid \right)
\end{equation}

\paragraph{A parabolic limit-state function}
A. Der Kiureghian and T. Dakessian \cite{der1998multiple} proposed this parabolic limit-state function whose equation is in the standard space:
\begin{equation}
g : \mathbf{x} \in \mathbb{R}^2 \longmapsto b - x_2 - \kappa (x_1 - \varepsilon )^2
\end{equation}
with $b = 5$, $\kappa = 0,5$ and $\varepsilon = 0.1$

\paragraph{A concave failure domain}
This function from R. Rackwitz \cite{rackwitz2001reliability} defines a concave failure domain in $\R^d$, $d$ being the dimension of the input space. It is defined in a lognormal input space ($d$ independent random variables with mean $\mu = 1$ and standard deviation $\sigma$) and thus normal-lognormal transformation is done before each call to the limit-state function. The original function in the lognormal input space writes as follows:
\begin{equation}
g : \mathbf{x} \in \mathbb{R}^d \longmapsto d + a \sigma \sqrt{d} - \sum \limits_{i=1}^d x_i
\end{equation}
with $a = 3$ and $\sigma = 0.2$, which becomes here:
\begin{equation}
g : \mathbf{u} \in \mathbb{R}^d \longmapsto d + a \sigma \sqrt{d} - \sum \limits_{i=1}^d \exp \left(- \dfrac{1}{2}\log(1+\sigma^2) + u_i\sqrt{\log(1+\sigma^2)} \right)
\end{equation}
Since the function is defined for any $d \in \N^*$ we try it with $d = 2$, $d=20$ and $d=50$ to study the behaviour of the algorithm depending on the dimension of the input space.

\begin{figure}[!htbp] \centering
\hspace{-0.7cm}
\parbox{0.3\textwidth}{\subfloat[First DoE]{\includegraphics[width=0.3\textwidth]{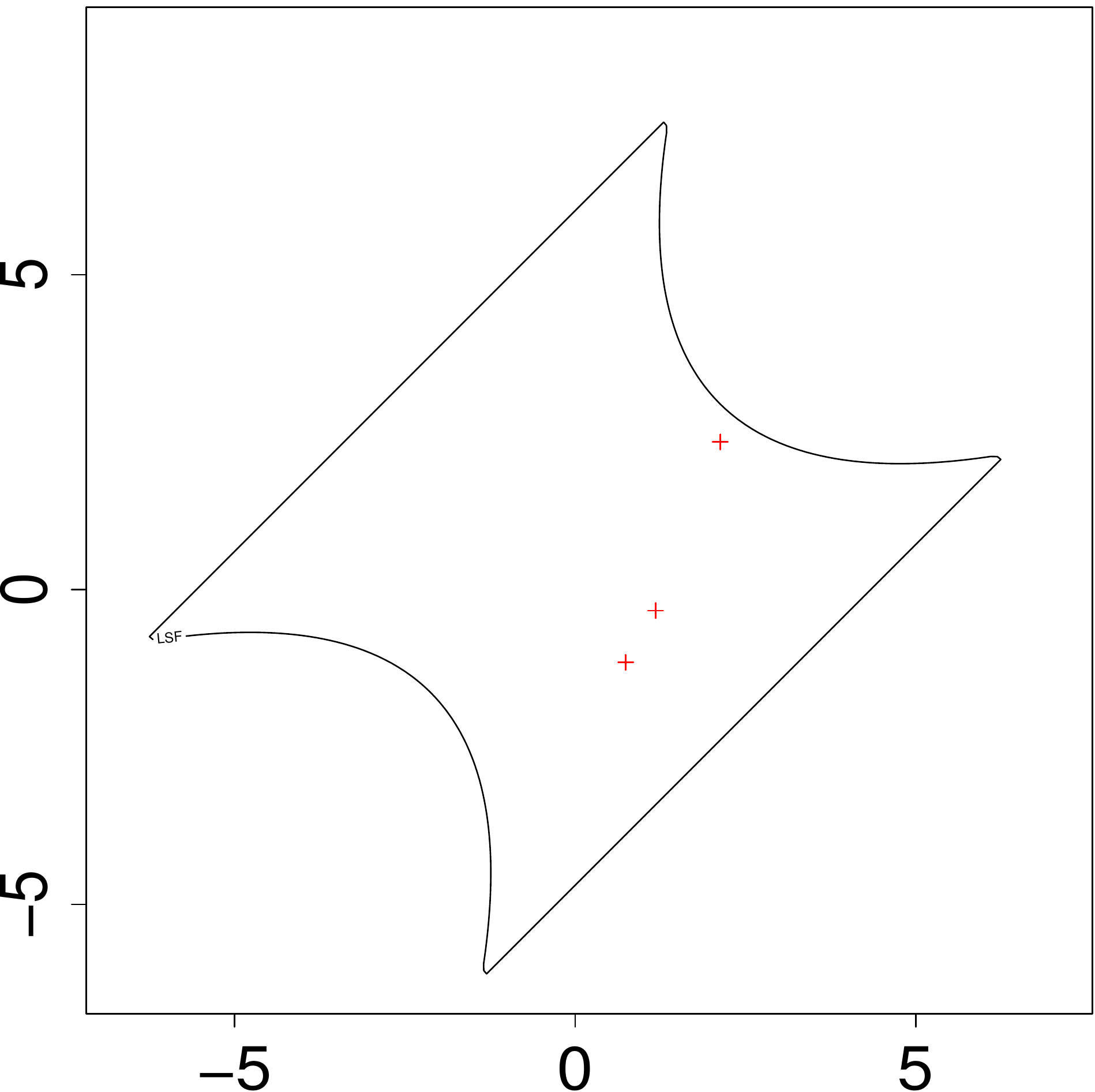}}}
\parbox{0.3\textwidth}{\subfloat[First Chain]{\includegraphics[width=0.3\textwidth]{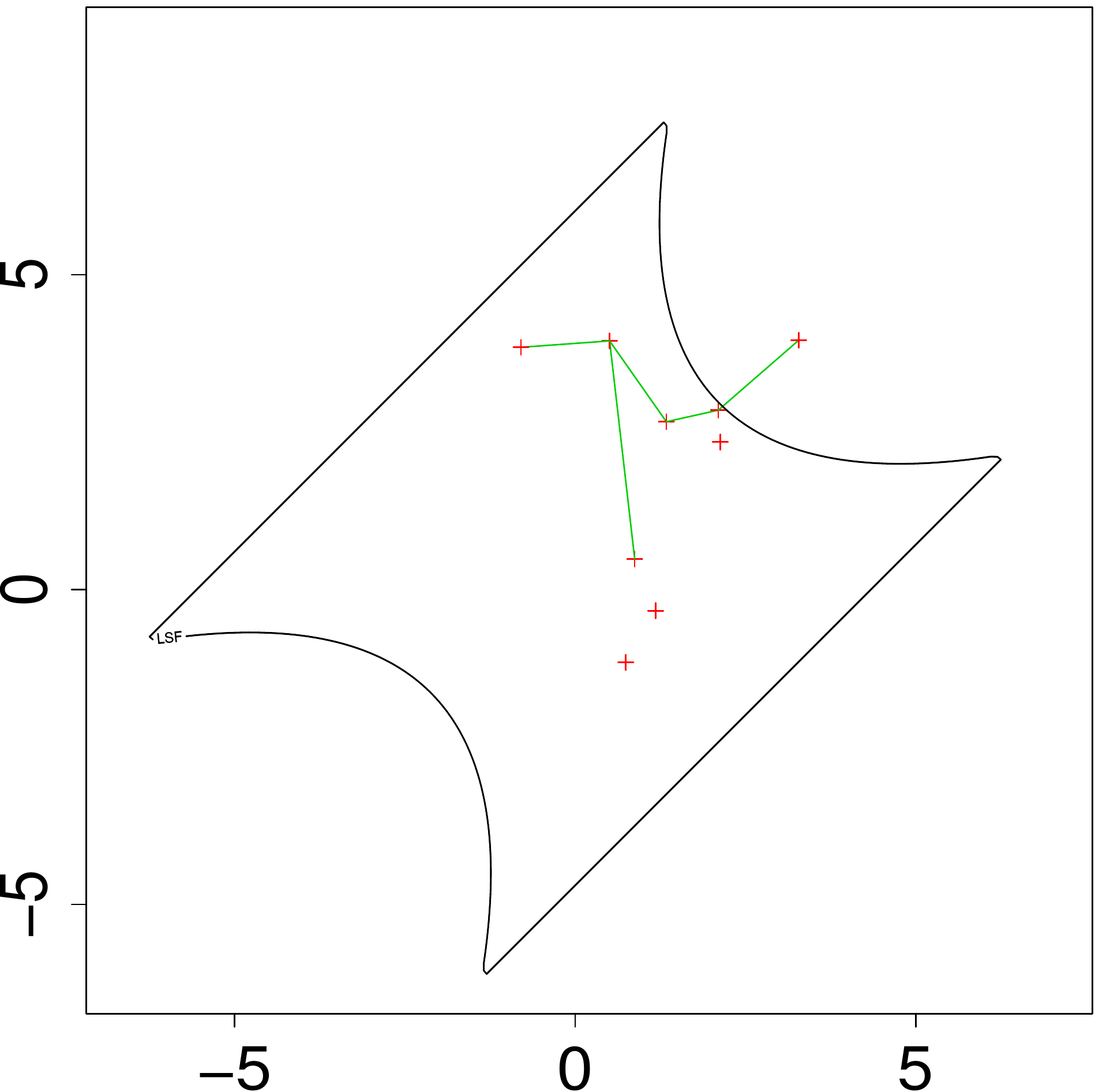}}}
\parbox{0.3\textwidth}{\subfloat[Final DoE and meta-model]{\includegraphics[width=0.3\textwidth]{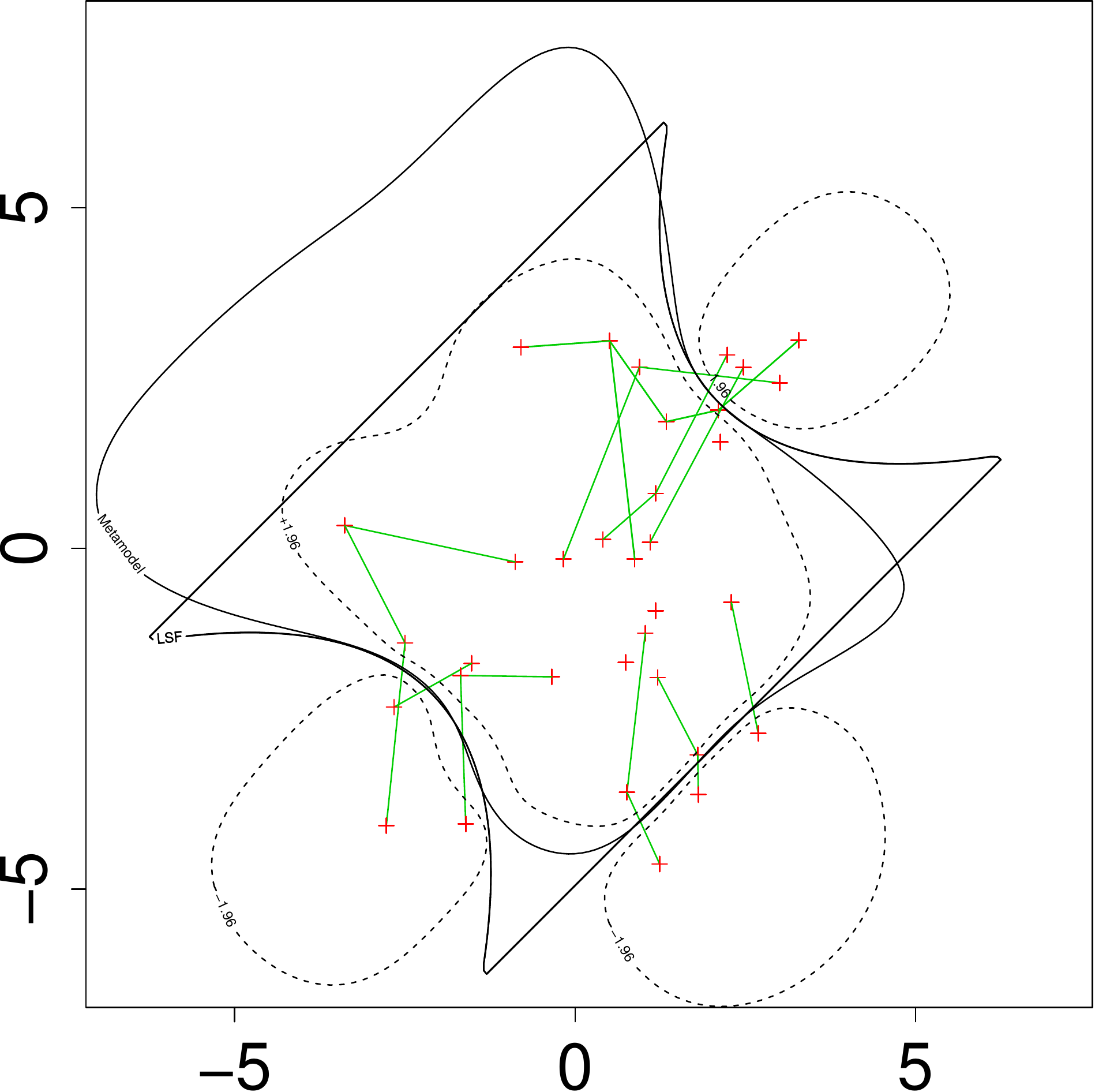}}}
\caption{Three steps of algorithm \ref{algo_MAPES} on a serial function}
\label{kiu_MAPES}
\end{figure}

Figure \ref{kiu_MAPES} shows three steps of Algorithm \ref{algo_MAPES}: the first DoE, the first Markov Chain and the final DoE with the meta-model. We notice that the second move took a wrong direction and proposed $\X_3$ was \emph{farther} from the failure domain. In this case, it started back from $\X_2$ while keeping this information for the train of the meta-model.

\paragraph{Results}
In this part we intend to compare the effective number of calls to the one given by the equation \eqref{Ncall_MAPES}. To compute the theoretical value, we use an estimate of the failure probability obtained with one run of algorithm \ref{algo_estim_proba} with $N=5000$ particles; it is indeed not necessary to get highly precise probability estimates as we focus on consistency between theory and practice only. We also present the number of iterations per particle to see how the progressive learning of the function modifies the behaviour of the algorithm.

Both results are summed up in Table \ref{MAPES_table} (adequacy to the equation \eqref{Ncall_MAPES}) and Table \ref{MAPES_table_iter} (evolution of the number of iterations per particle).
\begin{table}[!h]
\centering
\renewcommand{\arraystretch}{1}
\begin{tabular}{|c|c|c|c|c|c|c|}
\hline
Fonction & dim & Probability & Theor. $N_\text{call}$ & Pract. $N_\text{call}$ & Theor. $N_\text{iter}$ & Pract. $N_\text{iter}$ \\
\hline
Waarts & 2 & $2.275 \, 10^{-3}$ & 64.25 & 34 & 5.83 & 3.1 \\
Kiureghian & 2 & $2.946 \, 10^{-3}$ & 61.27 & 31 & 6.09 & 3 \\
Concave & 2 & $4.821 \, 10^{-3}$ & 56.35 & 36 & 5.35 & 3.3 \\
Concave & 20 & $2.273 \, 10^{-3}$ & 81.87 & 53 & 6.09 & 3.2 \\
Concave & 50 & $1.861 \, 10^{-3}$ & 113.87 & 82 & 6.29 & 3.1 \\
Oscillator 15 & 8 & $4.802 \, 10^{-3}$ & 62.75 & 48 & 5.38 & 3.9 \\
Oscillator 21.5 & 8 & $4.46 \, 10^{-5}$ & 109.75 & 78 & 10.02 & 6.9 \\
Oscillator 27.5 & 8 & $3.76 \, 10^{-7}$ & 156.94 & 113 & 14.79 & 10.4\\
Watermarking & 20 & $4.704 \, 10^{-11}$ & 258.80 & 259 & 23.78 & 23.8 \\
\hline
\end{tabular}
\caption{Getting into the failure domain with algorithm \ref{algo_MAPES}}
\label{MAPES_table}
\end{table}

\begin{table}[!h]
\centering
\renewcommand{\arraystretch}{1}
\begin{tabular}{|c|c|c|}
\hline
Fonction & Theor. $N_\text{iter}$ & Pract. $N_\text{iter}$ \\
\hline
Waarts & 5.83 & \{3 ; 4 ; 4 ; 3 ; 3 ; 2 ; 3 ; 3 ; 3 ; 3 \} \\
Kiureghian & 6.09 & \{4 ; 3 ; 2 ; 3 ; 4 ; 4 ; 3 ; 2 ; 3 ; 2 \} \\
Concave & 5.35 & \{2 ; 3 ; 5 ; 4 ; 3 ; 4 ; 3 ; 4 ; 3 ; 2 \} \\
Concave & 6.09 & \{3 ; 2 ; 4 ; 3 ; 4 ; 3 ; 4 ; 3 ; 4 ; 2 \} \\
Concave & 6.29 & \{4 ; 2 ; 4 ; 3 ; 3 ; 3 ; 3 ; 5 ; 2 ; 2 \} \\
Oscillator 15 & 5.38 & \{7 ; 2 ; 3 ; 2 ; 5 ; 4 ; 4 ; 5 ; 2 ; 5 \} \\
Oscillator 21.5 & 10.02 & \{7 ; 6 ; 5 ; 9 ; 5 ; 11 ; 6 ; 7 ; 7 ; 6 \} \\
Oscillator 27.5 & 14.79 & \{15 ; 9 ; 8 ; 7 ; 10 ; 15 ; 9 ; 15 ; 6 ; 10 \} \\
Watermarking & 23.78 & \{34 ; 61 ; 16 ; 25 ; 16 ; 14 ; 13 ; 18 ; 20 ; 21 \} \\
\hline
\end{tabular}
\caption{Number of iterations per particle in algorithm \ref{algo_MAPES}}
\label{MAPES_table_iter}
\end{table}
Finally practical results show a good behaviour of the algorithm on different situations and it appears to be quite versatile. Apart from specific Watermarking test-case the number of iterations  does not seem to depend on the rank of the failing point. In this latter case the algorithm struggled to get the two first failing points and then went a lot faster into the failure domain: it is to suppose that it found a specific path and then put all the points in the same \emph{area}.

\subsection{Perspectives}

The adaptation of Algorithm \ref{algo_atteinte_defaillance} to the construction of first DoE seems to be a good mean to get into the failure domain. Once \emph{there}, two questions remain open:
\begin{itemize}
\item how to prevent the algorithm from moving the particles all at the same place? in other words: how to make the meta-model \emph{understand} that it is not necessary to go back to regions it has already visited? Self-avoiding walks could be used here.
\item once in the failure domain, what should be an \emph{efficient} strategy to refine the meta-model? Indeed usual refinement strategies appear here like a backtracking as there are all global strategies and would not use the local accuracy of the meta-model to the limit-state function close to the failing point. Some strategies trying to follow the boundary from the failing points seem to show promising results.
\end{itemize} 

\appendix
\section*{Appendix}
\Closesolutionfile{ann}
\input{demo}

\newpage

\bibliographystyle{abbrv}
\bibliography{biblio}

\end{document}